\documentclass[reprint, amssymb, amsmath, aip,cha, groupedaddress,frontmatterverbose,]{revtex4-1}  

\usepackage{graphicx}
\usepackage{dsfont}
\usepackage{amsmath,amssymb,amsfonts,mathtools}
\usepackage{ulem}
\usepackage{color}
\usepackage{multirow}
\usepackage{float}
\usepackage{soul}
\newcommand*{\citen}[1]{%
  \begingroup
    \romannumeral-`\x 
    \setcitestyle{numbers}%
    \cite{#1}%
  \endgroup   
}

\begin{document}

\title{Symmetric Quasi Classical Dynamics with Quasi Diabatic Propagation Scheme}
\author{Juan Sebasti\'an Sandoval C.}%
\author{Arkajit Mandal}%
\author{Pengfei Huo}%
\email{pengfei.huo@rochester.edu}
\affiliation{Department of Chemistry, University of Rochester, 120 Trustee Road, Rochester, New York 14627, United States}%

\begin{abstract}
We apply a recently developed quasi-diabatic (QD) scheme to the symmetric quasi-classical (SQC) approach for accurate quantum dynamics propagation. By using the adiabatic states as the quasi-diabatic states during a short-time quantum dynamics propagation, the QD scheme allows directly interfacing diabatic SQC method with commonly used adiabatic electronic structure calculations, thus alleviate any non-trivial theoretical efforts to reformulate SQC in the adiabatic representation. Further, the QD scheme ensures a stable propagation of the dynamics and allows using a much larger time step compared to directly propagating SQC dynamics in the adiabatic representation. This is due to the fact that the QD scheme does not explicitly require non-adiabatic couplings that could exhibit highly peaked values during non-adiabatic dynamics propagation. We perform the QD-SQC calculations with a wide range of model non-adiabatic systems to demonstrate the accuracy of the proposed scheme. This study opens up the possibility for combining accurate diabatic quantum dynamics methods such as SQC with any adiabatic electronic structure calculations for non-adiabatic on-the-fly propagations.
\end{abstract}
\maketitle
\section{Introduction}
The recently developed symmetric quasi-classical (SQC) approach\cite{MillerJCP13,SCFaradayMiller} has shown great promise for providing accurate non-adiabatic dynamics.\cite{Miller2014ET,MillerJCTC16,CottonJCP16,TaoSF} By using the window function as the population estimator, SQC can significantly reduce the number of trajectories required for convergence,\cite{SCFaradayMiller} while at the same time, recovers detailed balance with a reasonable accuracy \cite{MillerJCP15,Subotnik2016Equilibrium} and provide a full description of the electronic density matrix.\cite{MillerJCP16,CottonJCP16} New developments based on this scheme, such as coherence-controlled SQC\cite{Tao2016-2,Tao2016-3} or extended SQC\cite{Geva2018} have further improved the accuracy of this approach. That being said, SQC still faces some intrinsic deficiencies, such as inverted potential\cite{Subotnik2016Equilibrium} that impact its numerical performance in recovering exact thermal equilibrium populations, or failed to achieve convergence for describing vibrational relaxation process.\cite{Subotnik2018-VER} Nevertheless, the quasi-classical nature of the SQC dynamics, together with many appealing features mentioned above make it a promising method to simulate non-adiabatic on-the-fly dynamics of complex molecular systems, \cite{SCFaradayMiller} providing an attractive alternative that departs from the commonly used fewest-switches surface hopping (FSSH) approach.\cite{Tully}

When performing on-the-fly simulations, the {\it adiabatic} representation is convenient for electronic structure calculations. Thus, the typical strategy for applying SQC (or other recently developed diabatic dynamics methods\cite{ananth2007,HuoJCP2012,Hsieh13mol}) to ``real'' molecular systems is to reformulate them in the {\it adiabatic} representation,\cite{MillerAdiabatic} which usually requires additional non-trivial theoretical efforts. Moreover, the adiabatic version of these methods are computationally inconvenient due to the presence of the first and second-order derivative couplings,\cite{MillerAdiabatic} which could potentially lead to numerical instabilities during dynamical propagations. Recently developed kinematic momentum (KM)-SQC approach\cite{MillerAdiabatic} uses the \textit{kinematic} momentum instead of the canonical momentum as the dynamical variable, and explicitly eliminates the presence of the second-derivative coupling inside the equation of motion, thus significantly reducing the numerical cost. However this approach as well as other adiabatic mapping approaches\cite{ananth2007,HuoJCP2012,Hsieh13mol} do require computing the time-dependent non-adiabatic coupling, therefore, it might encounter numerical instabilities when these couplings are highly peaked. 

An alternative route is to employ {\it diabatic} electronic structure approaches\cite{DiabaticARPC,SubotnikDIA,Blumberger,Yang2012,Hammes-Schiffer2011DIA} or diabatization procedures\cite{Ananth:2017,Kretchmer:2013} to construct globally well-defined diabatic states. Under this representation, the derivative couplings explicitly vanish, providing a convenient representation for developing various dynamics approaches and propagating quantum dynamics.  However, these diabatic based electronic structure approaches are not routinely available despite recent theoretical progress.\cite{DiabaticARPC,SubotnikDIA} Further, the diabatization procedures that construct globally defined diabatic models by fitting the adiabatic surfaces might introduce additional error. At this point, it almost seems that (1) if we want to use diabatic-based dynamics approach, we need to construct globally well-defined diabatic states for the system, (2) if we want to use adiabatic energies and gradients to perform on-the-fly simulation, we have to use dynamics approaches that are explicitly formulated in the adiabatic representation.\cite{MillerAdiabatic,ananth2007,HuoJCP2012,Hsieh13mol} 

However, we realize\cite{Huo2018} that in order to use diabatic approaches for quantum dynamics propagation, we do not actually need a globally well-defined diabatic surface; rather, we only need a set of locally defined diabatic states. To this end, we have developed the quasi-diabatic (QD) propagation scheme\cite{Huo2018} which uses the adiabatic states as the quasi-diabatic states (local diabatic states) during a short-time propagation, and dynamically update the QD states between two consecutive short-time propagations. This propagation scheme explicitly addresses the {\it discrepancy} between accurate {\it diabatic} quantum dynamics approaches and routinely available {\it adiabatic} electronic structure methods, allowing a seamless interface between them without any additional non-trivial efforts.

In this work, we apply the QD propagation scheme\cite{Huo2018} with the {\it diabatic} SQC approach for non-adiabatic dynamics simulations. We refer to this approach as QD-SQC throughout this work. By using the QD scheme, we avoid any additional non-trivial effort to reformulate the diabatic SQC approach back into the adiabatic representation. Further, by avoiding the explicit presence of non-adiabatic couplings, the QD scheme provides a more robust approach for dynamical propagation compared to the other adiabatic schemes. We demonstrate the accuracy and the stability of QD-SQC dynamics with a variety of non-adiabatic models. For model calculations with the strict diabatic models, QD-SQC provides the exactly the same results as obtained from the diabatic SQC, and provide a robust propagation regardless of the presence of highly peaked or diverging non-adiabatic couplings. We further demonstrate the applicability and accuracy of QD-SQC by using adiabatic vibronic states of a coupled proton-electron model as the QD states,\cite{Hazra:2010,Hazra:2011} where obvious low-dimensional diabatic vibronic states are not readily available, without non-trivial diabatization schemes. This study opens up the possibility for using QD-SQC to perform accurate on-the-fly non-adiabatic quantum dynamics for realistic and complex molecular systems.

\section{Theory and Method}
\subsection{MMST Mapping Hamiltonian}\label{sec:mmst}
We begin with a brief outline of the Meyer-Miller-Stock-Thoss mapping Hamiltonian,\cite{MMJCP79,StockThossPRL97,StockThossPRA99} which is one of the basic ingredients for many non-adiabatic dynamics approaches. The total Hamiltonian for a given molecular system can be expressed as a sum of the nuclear kinetic energy operator $\hat{T}$ and the electronic Hamiltonian operator $\hat{V}(\hat{\bf r},\hat {\bf R})$ as follows
\begin{equation}\label{eqn:totalH}
\hat {H} = \hat {T} + \hat{V}(\hat{\bf r},\hat {\bf R}). 
\end{equation}
Here, $\hat{\bf r}$ represents the coordinate operator of the electronic degrees of freedom (DOF), and $\hat{\bf R}$ represents the coordinate operator of the nuclear DOF. 

We start by expressing the total Hamiltonian in Eqn.~\ref{eqn:totalH} with strict {\it diabatic} basis $\{|i\rangle,|j\rangle\}$, {\it i.e.}, a set of basis that does not explicitly depend on nuclear positions. With the diabatic basis, the total Hamiltonian is expressed as follows
\begin{equation}
\hat H = \hat T + \sum_{ij} V_{ij}(\hat{\bf R})|i \rangle\langle j|, 
 \end{equation}
where $V_{ij}(\hat{\bf R})=\langle i|\hat{V}(\hat{\bf r},\hat {\bf R})|j\rangle$ is the state-dependent potential. By using the mapping representation of Meyer-Miller-Stock-Thoss\cite{MMJCP79,StockThossPRL97,StockThossPRA99} to transform the discrete electronic states into continuous variables, we have
\begin{equation} \label{eq:mmst}
|i\rangle\langle j| \rightarrow  {\hat a}_{i}^\dagger {\hat a}_{j},
\end{equation} 
where ${\hat a}^\dagger_{i} =({\hat q}_{i} - i{\hat p}_{i})/\sqrt{2}$ and ${\hat a}_{j} =({\hat q}_{j} + i{\hat p}_{j})/\sqrt{2}$. With this transformation, the original {\it diabatic} Hamiltonian is transformed into the following MMST mapping Hamiltonian
\begin{equation} \label{eq:map} 
\hat{H}_\mathrm{m}=\hat{T}+{1\over2}\sum_{ij}V_{ij}(\hat{R})\left(\hat{p}_{i}\hat{p}_{j}+\hat{q}_{i}\hat{q}_{j}-2\gamma\delta_{ij}\right), 
\end{equation}
where $\gamma=0.5$ is the zero-point energy (ZPE) for the mapping harmonic oscillators (historically, it is recognized as the Langer correction by Meyer and Miller\cite{MMJCP79} for the quasi-classical description). Up to here, there is no approximation. 

Instead of solving the equation of motion quantum mechanically, SQC assumes that the coupled electronic-nuclear dynamics is governed by the following Hamiltonian\cite{SCFaradayMiller} 
\begin{equation} \label{eq:mapham} 
{H}_\mathrm{m}={{{\bf P}^2}\over {2M}}+{1\over2}\sum_{ij}V_{ij}(R)\left(p_{i}p_{j}+q_{i}q_{j}-2\gamma\delta_{ij}\right).
\end{equation}
Classical trajectories are generated based on the following Hamilton's equations of motion
\begin{eqnarray} \label{eq:mapeqn} 
\dot q_{i} &=& \partial H_\mathrm{m}/ \partial p_{i};~~\dot p_{i} = -\partial H_\mathrm{m} / \partial q_{i}\\
\dot {\bf R} &=& \partial H_\mathrm{m}/ \partial {\bf P};~~ \dot {\bf P}=-\partial H_\mathrm{m} / \partial {\bf R}= {\bf F}, \nonumber
\end{eqnarray}
with the nuclear force expressed as
\begin{equation}\label{eqn:nucforce}
{\bf F}=-{1\over 2}\sum_{ij}\nabla V_{ij}({\bf R})\bigg(p_{i}p_{j}+q_{i}q_{j}-2\gamma\delta_{ij}\bigg).
\end{equation}
Thus, MMST mapping Hamiltonian provides a consistent classical footing for both electronic and nuclear DOFs. The non-adiabatic transitions among electronic states are mapped onto the classical motion of fictitious harmonic oscillators.

\subsection{Symmetric Window Function Estimator}\label{sec:sqc}
The equation of motion generated from $\hat{H}_\mathrm{m}$ (Eqn.~\ref{eq:mapeqn}) is equivalent to the Enhrefest dynamics.\cite{ananth2007} However, by using a window function\cite{MillerJCP13} to restrain the initial mapping conditions and estimate the time-dependent population, SQC approach can significantly improve the numerical performance of non-adiabatic dynamics calculations, even with a Ehrenfest type equation of motion.

The SQC window function is formulated with the action-angle variables, $\{n_{i}, \theta_{i}\}$, which are related to the canonical mapping variables through the following relations
\begin{equation}
n_i = \frac{1}{2}\left(p_i^2 + q_i^2 - 2\gamma \right);~~~\theta_i =-\tan^{-1}\left( \frac{p_i}{q_i}\right), 
\end{equation}
and the inverse relations
\begin{equation}
q_i = \sqrt{2 ( n_i + \gamma) }\cos(\theta_i);~~p_i =-\sqrt{2 ( n_i + \gamma )}\sin(\theta_i).
\end{equation}

The SQC approach can be viewed as the classical Wigner model of the action-angle mapping variables\cite{SCFaradayMiller,MillerJCP16}  for computing population dynamics with the following expression
\begin{eqnarray}\label{eqn:wignersqc}
&&\rho_{jj}(t)=\mathrm{Tr}_{\bf R}\left[\hat{\rho}_{\bf R}|i\rangle\langle i|e^{i\hat{H}t/\hbar}|j\rangle\langle j|e^{-i\hat{H}t/\hbar}\right] \\
&&\approx\frac{1}{\left(2\pi \hbar\right)^{N+M}}\int d\boldsymbol{\tau} W_{\bf R}(\mathbf{P},\mathbf{R})W_i(\mathbf{n}(0))W_j(\mathbf{n}(t)).\nonumber
\end{eqnarray}
Here, $\hat{\rho}(0)=|i\rangle \langle i|\otimes \hat{\rho}_{\bf R}$ is the initial density operator, $W_{\bf R}(\mathbf{P},\mathbf{R})$ is the Wigner density of $\hat{\rho}_{\bf R}$ that contains $M$ total nuclear DOF, $\mathbf{n} = \{n_1,n_2,...n_{N}\}$ is the action variable vector for $N$ electronic states, with the corresponding angle variable vector $\boldsymbol{\theta} = \{\theta_1,\theta_2,...\theta_{N}\}$, and $d\boldsymbol{\tau}\equiv d\mathbf{P} d\mathbf{R} d\mathbf{n} d\boldsymbol{\theta}$. Further, $W_i(\mathbf{n})$ is the Wigner transformed action variables\cite{MillerJCP16} 
\begin{equation}\label{eqn:window}
W_i (\mathbf{n}) = \delta (n_i - 1)\prod_{i \neq j} \delta(n_j),     
\end{equation}
The above results can be viewed as the Bohr-Sommerfeld quantization rule.\cite{MillerJCP16}  

In the SQC approach where the classical dynamics is used to solve Eqn.~\ref{eq:mapeqn}-\ref{eqn:nucforce}, the delta functions are better to be broadened by ``pre-limit" delta functions, {\it i.e.}, the \textit{window functions} that center at the integer values of the initial and final action variables, in order to facilitate the numerical convergence.\cite{SCFaradayMiller} Further, in SQC approach,  $\gamma$ in Eqn.~\ref{eq:mapham} is viewed as a parameter\cite{MillerJCP13,StockConical} instead of the ZPE of the mapping oscillator (with a value of 0.5). 

Because these window functions are viewed as pre-limit delta function, {\it i.e.}, an approximation of Eqn.~\ref{eqn:window}, they do not have a unique form,\cite{CottonJCP16} and thus allowing the engineering aspect of the SQC approach. One can choose the following square window function\cite{MillerJCP13}
\begin{equation}\label{eqn:square}
W_{i}(t) = w_1(n_i) \prod_{j\neq i} w_0(n_j),
\end{equation}
where $w_n$ is the square window function expressed as follows
\begin{equation}
w_n (n_j)=\frac{1}{2 \gamma} h\left(\gamma - |n_j-n|\right).
\end{equation}
Here, $h(z)$ is the Heaviside function and $n$ (either 0 or 1) is the electronic quantum number. 

\begin{figure}
 \centering
  \begin{minipage}[h]{\linewidth}
     \centering
     \includegraphics[width=\linewidth]{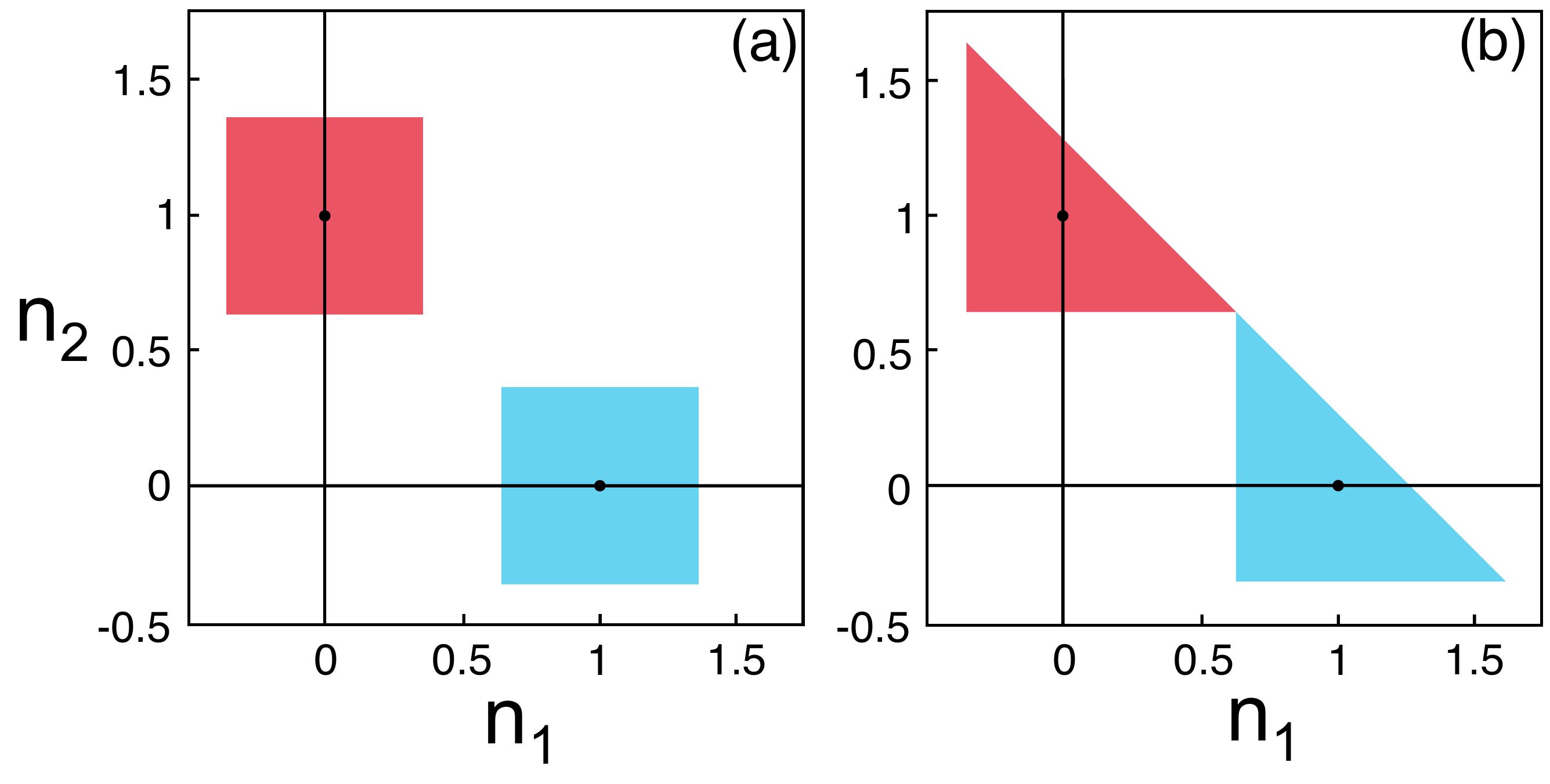}
  \end{minipage}%
   \caption{Two possile choices for the window function in the action space: (a) square histogram windows with $\gamma ={1\over2}(\sqrt{3}-1)\approx 0.366$ and (b) triangle histogram windows with $\gamma = {1\over3}$. Here, the blue windows are used to estimate the population of state 1, and the red windows are used to estimate the population of state 2.}
\label{fig:sqc}
\end{figure}

Fig.~\ref{fig:sqc}a depicts the above window function for a system with two electronic states, with the width $\gamma ={1\over2}(\sqrt{3}-1)\approx 0.366$ suggested by Cotton and Miller.\cite{MillerJCP13} Numerical results obtained from this window function have shown excellent agreement with the exact quantum dynamics for various model non-adiabatic systems.\cite{MillerJCTC16,SCFaradayMiller}

Fig.~\ref{fig:sqc}.b illustrates a recently proposed triangle window function.\cite{CottonJCP16} For two-level systems, this triangle windowing function can be described with the following expression
\begin{eqnarray}\label{eqn:triangle}
    W_1(t)&=&2 h(n_1+\gamma-1) h(n_2+\gamma)h(2-2\gamma-n_1-n_2)\nonumber\\
    \\
    W_2(t)&=&2 h(n_1+\gamma)h(n_2+\gamma-1)h(2-2\gamma-n_1-n_2), \nonumber
\end{eqnarray}
with the width $\gamma = {1\over3}$. This window function has shown a consistently better performance for two-level systems \cite{SCFaradayMiller,CottonJCP16} compared to the square window function, as well as a more accurate description for the non-adiabatic transition rate over a broad range of electronic couplings.\cite{CottonJCP16} 

The time-dependent population at time $t$ is then calculated by applying the window function estimator to actions variables $\{n_j (t)\}$
for an ensemble of trajectories. Starting from the initial diabatic state $|i \rangle$, the time-dependent population of the states $|j \rangle$ is computed with Eqn.~\ref{eqn:wignersqc}. However, by using the window function estimator, the total population is no longer properly normalized due to the fraction of trajectories that move out of any given window.\cite{MillerJCP13} Thus, the population has to be normalized\cite{MillerJCP13} with the following procedure
\begin{equation}\label{eq:popNormalize}
{{\rho}_{jj}(t)}/{\sum_{k=1}^N {\rho}_{kk}(t)}\rightarrow {\rho}_{jj}(t).
\end{equation}
It should be noted that SQC is different compared to ``Ehrenfest dynamics", despite that they use the same equation of motion for the coupled electronic-nuclear DOFs.\cite{SCFaradayMiller,Subotnik2016Equilibrium} The boundary conditions enforced by the window functions in SQC helps to eliminate several well-known deficiencies in Ehrenfest dynamics, such as the breakdown of detailed balance.\cite{DetailedBalanceMQC,MillerJCP15,Subotnik2016Equilibrium} 

Despite its simplicity, SQC has shown accurate description for non-adiabatic dynamics in a broad range of model systems.\cite{Miller2014ET,MillerJCTC16,CottonJCP16,TaoSF} It can also recover detailed balance with reasonable accuracy \cite{MillerJCP15,Subotnik2016Equilibrium} and reach convergence with just a few thousand of trajectories. \cite{SCFaradayMiller,MillerJCP13,MillerJCTC16} It thus show great promise to accurately and efficiently perform {\it ab-initio} on-the-fly simulations for molecular systems.

\subsection{Quasi-Diabatic (QD) Propagation Scheme}\label{sec:QD}
For real molecular systems, strict diabatic states $\{|i\rangle,|j\rangle\}$ are neither uniquely defined nor routinely available, despite recent theoretical progress.\cite{DiabaticARPC,SubotnikDIA,Blumberger,Yang2012,Hammes-Schiffer2011DIA} Rather, it is convenient to solve the electronic structure problem under the {\it adiabatic} representation with the following eigenequation 
\begin{equation}\label{eq:eigenvalue}
\hat{V}(\hat{\bf r}, {\bf R}) |\Phi_\alpha({\bf R})\rangle = E_{\alpha}(\mathbf{R})|\Phi_\alpha({\bf R})\rangle.
\end{equation}
Here, $\hat{V}(\hat{\bf r}; {\bf R})$ is the electronic Hamiltonian operator defined in Eqn.~\ref{eqn:totalH} at a given nuclear configuration ${\bf R}$, and $|\Phi_{\alpha}({\bf R})\rangle$ is the {\it adiabatic} state, {\it i.e.}, the eigenstate of $\hat{V}(\hat{\bf r}; {\bf R})$, with the corresponding eigenvalue $E_{\alpha}(\mathbf{R})$.  Most of the commonly used electronic structure methods are based on solving the above equation, providing eigenenergies and eigenfunctions under this representation. 

The total Hamiltonian Eqn.~\ref{eqn:totalH}, on the other hand, has a rather complicated expression (see Appendix A) under the {\it adiabatic} representation. This is due to the fact that adiabatic states are not the eigenfunctions of the nuclear kinetic energy operator $\hat{T}$. It is more convenient to develop new quantum dynamics methods in the strict diabatic representation (such as the diabatic SQC introduced in Sec.~\ref{sec:mmst}-\ref{sec:sqc}). Thus, the typical strategy for applying new quantum dynamics approaches (like SQC) to ``real'' molecular systems is to reformulate them in the adiabatic representation.\cite{MillerAdiabatic} However, this reformulating process usually requires additional non-trivial theoretical efforts, \cite{ananth2007,MillerAdiabatic,HuoJCP2012,Hsieh13mol} and the resulting adiabatic version of these methods are computationally inconvenient due to the presence of the first and second-order derivative couplings,\cite{MillerAdiabatic} which could lead to numerical instabilities during dynamical propagations. 

To address this discrepancy between accurate quantum dynamics methods in the diabatic representation and the electronic structure methods in the adiabatic representation, we have developed quasi-diabatic (QD) propagation scheme.\cite{Huo2018} Here, we briefly summarize it by considering a short-time propagation of the nuclear DOF during $t\in[t_1, t_2]$, where the nuclear positions evolve from ${\bf R}(t_1)$ to ${\bf R}(t_2)$, and the corresponding adiabatic states are $\{|\Phi_{\alpha}({\bf R}(t_1))\rangle\}$ and $\{|\Phi_{\mu}({\bf R}(t_2))\rangle\}$. 

{\it The essential idea of the QD scheme} is to use the nuclear geometry at time $t_1$ as the reference geometry, ${\bf R_{0}}\equiv {\bf R}(t_1)$, and use the adiabatic basis $\{|\Phi_{\alpha}({\bf R}(t_1))\rangle\}$ as the {\it quasi-diabatic} basis during this short-time quantum dynamics propagation, such that
\begin{equation}
|\Phi_{\alpha}({\bf R_{0}})\rangle\equiv|\Phi_{\alpha}({\bf R}(t_1))\rangle,~~\mathrm{for}~t\in[t_1,t_2].
\end{equation}
With the above QD basis (often called the ``crude adiabatic basis"), the derivative couplings vanish in a trivial way, and $\hat{V}(\hat{\bf r};{\bf R})$ has off-diagonal elements. Because the electronic wavefunction changes rapidly with the motion of the nuclei, the QD basis is only convenient when the nuclear geometry $\bf R$ is close to the reference geometry ${\bf R_0}$. Thus, during the next short-time propagation segment $t\in[t_2,t_3]$, we choose to use a new reference geometry ${\bf R'_{0}}\equiv {\bf R}(t_2)$ and {\it quasi-diabatic} basis $|\Phi_{\mu}'({\bf R'_{0}})\rangle\equiv|\Phi_{\mu}({\bf R}(t_2))\rangle$. 

We emphasize that there is always a non-removable part of the derivative coupling over the {\it entire configurational space} for polyatomic systems.\cite{Mead:1982}  This is a well-known result in literature.\cite{DiabaticARPC,SubotnikDIA} Here, the QD scheme circumvents this challenge by only requiring a set of {\it locally-defined diabatic} states, such that the derivative couplings vanish within the {\it configurational subspace} during a given short-time propagation.

Compared to the adiabatic representation, the advantage of the QD basis is that all of the derivative couplings vanish. As a consequence, the total Hamiltonian operator and the corresponding quantum dynamics propagation adapt a simpler form in the QD representation. With the nuclear geometry close to the reference geometry in each step, the QD states remains to be a convenient and compact basis in each short-term propagation segment. In addition, because of the diabatic nature of the QD basis, one can use any diabatic based approach to propagate the quantum dynamics. These approaches usually require diabatic energies, electronic couplings, and nuclear gradients. Between $[t_1,t_2]$ propagation and $[t_2,t_3]$ propagation segments, all of these quantities will be transformed from $\{|\Phi_{\alpha}({\bf R_{0}})\rangle\}$ basis to $\{|\Phi'_{\mu}({\bf R'_{0}})\rangle\}$ basis.

With the above idea in mind, it is straightforward to obtain electronic couplings and nuclear gradients in the QD basis. During the $t\in[t_1,t_2]$ short-time propagation, the electronic Hamiltonian operator $\hat V (\hat {\bf r}; {\bf R}(t))$ is evaluated under the QD states as
 \begin{equation}\label{eqn:vijt}
 V_{\alpha\beta}({\bf R}(t))  = \langle \Phi_\alpha ({\bf R_0})| \hat V (\hat{\bf r};{\bf R}(t))|  \Phi_\beta({\bf R_0})\rangle.
 \end{equation}
In practical on-the-fly calculations, the above quantity can be obtained from a linear interpolation between $V_{\alpha\beta}({\bf R}(t_{1}))$ and $V_{\alpha\beta}({\bf R}(t_{2}))$ as follows\cite{Rossky-Webster}
\begin{equation}\label{eqn:interpolation}
V_{\alpha\beta}({\bf R}(t)) = V_{\alpha\beta}(\mathbf{R}(t_{1}))+\frac {(t - t_{1})}{(t_{2} - t_{1})}\bigg[V_{\alpha\beta}(\mathbf{R}(t_{2})) - V_{\alpha\beta}(\mathbf{R}(t_{1}))\bigg].
\end{equation} 
Here, the matrix elements $V_{\alpha\beta}({\bf R}(t_{1})) =\langle \Phi_\alpha ({\bf R_0})| \hat V (\hat{\bf r};{\bf R}(t_1))| \Phi_\beta ({\bf R_0})\rangle = E_{\alpha}({\bf R}(t_1))\delta_{\alpha\beta}$, and the matrix elements $V_{\alpha\beta}({\bf R}(t_{2}))$ can be easily computed as follows
\begin{equation}\label{eqn:elect2}
V_{\alpha\beta}({\bf R}(t_{2})) =\sum_{\mu\nu}b_{\alpha\mu}\langle \Phi_{\mu}({\bf R}(t_{2}))| \hat {V} (\hat{\bf r};{\bf R}(t_2))|\Phi_{\nu}({\bf R}(t_{2}))\rangle  b^{\dagger}_{\beta\nu},   
 \end{equation}
where $\langle \Phi_{\mu}({\bf R}(t_{2}))| \hat {V} (\hat{\bf r};{\bf R}(t_2))|\Phi_{\nu}({\bf R}(t_{2})) \rangle=E_{\mu}({\bf R}(t_2))\delta_{\mu\nu}$, $b_{\alpha\mu}= \langle \Phi_{\alpha}({\bf R_0})|\Phi_{\mu}({\bf R}(t_{2}))\rangle$, and $b^{\dagger}_{\beta\nu} = \langle \Phi_{\nu}({\bf R}(t_{2}))|\Phi_{\beta}({\bf R_0})\rangle$.

Similarly, the nuclear gradients on electronic Hamiltonian matrix elements $\nabla V_{\alpha\beta}({\bf R}(t_{2}))\equiv \partial V_{\alpha\beta}({\bf R}(t_{2}))/\partial {\bf R}$ are evaluated as
\begin{eqnarray}\label{eqn:nucgrad}
&&\nabla V_{\alpha\beta}({\bf R}(t_{2}))=\nabla \langle \Phi_\alpha({\bf R_{0}})| \hat V ({\hat{\bf r}};{\bf R}(t_2))| \Phi_\beta ({\bf R_{0}})\rangle \\
&&=\langle \Phi_\alpha ({\bf R_{0}})| \nabla \hat V ({\hat{\bf r}};{\bf R}(t_2))| \Phi_\beta ({\bf R_{0}})\rangle \nonumber \\
&&=\sum_{\mu\nu} b_{\alpha\mu}\langle \Phi_{\mu}({\bf R}(t_{2}))|\nabla \hat V (\hat{\bf r};\mathbf{R}(t_2))|\Phi_{\nu}({\bf R}(t_{2})) \rangle b^{\dagger}_{\beta\nu}. \nonumber
\end{eqnarray}  
Here, we have used the fact that $\{|\Phi_\alpha({\bf R_0})\rangle\}$ is a {\it diabatic} basis during the $[t_1, t_2]$ propagation, which allows moving the gradient operator to bypass $\langle \Phi_\alpha({\bf R_{0}})|$. Moreover, we have inserted the resolution of identity $\sum_{\mu} |\Phi_{\mu}({\bf R}(t_2))\rangle\langle \Phi_{\mu}({\bf R}(t_2))|=\mathds{1}$, where we explicitly assume that the QD basis at nuclear position ${\bf R}(t_2)$ is complete. We emphasize that Eqn.~\ref{eqn:nucgrad} includes derivatives with respect to all possible sources of the nuclear dependence, including those from the adiabatic potentials as well as the adiabatic orbitals. This can be simply verified by using the basic property of the adiabatic states $\langle\Phi_{\mu}({\bf R}(t_{2}))|\nabla\hat{V}(\hat{\bf r};\mathbf{R}(t_2))|\Phi_{\nu}({\bf R}(t_{2}))\rangle=\nabla V_{\mu\nu}(\mathbf{R}(t_2))+E_{\nu}(\mathbf{R}(t_2))\langle\Phi_{\mu}({\bf R}(t_{2}))| \nabla\Phi_{\nu}({\bf R}(t_{2}))\rangle +E_{\mu}(\mathbf{R}(t_2))\langle \nabla \Phi_{\mu}({\bf R}(t_{2}))| \Phi_{\nu}({\bf R}(t_{2}))\rangle$, where $V_{\mu\nu}(\mathbf{R}(t_2))=\langle \Phi_{\mu}({\bf R}(t_{2}))| \hat {V} (\hat{\bf r};{\bf R}(t_2))|\Phi_{\nu}({\bf R}(t_{2})) \rangle$. Plugging the above equality into the last line of Eqn.~\ref{eqn:nucgrad},  and noticing the resolution of identity $\sum_{\mu} |\Phi_{\mu}({\bf R}(t_2))\rangle\langle \Phi_{\mu}({\bf R}(t_2))|=\mathds{1}$, we can easily verify that the nuclear gradient in Eqn.~\ref{eqn:nucgrad} is equivalent to the following expression
\begin{eqnarray}\label{eqn:adgrad}
&&\sum_{\mu\nu} b_{\alpha\mu} \langle \Phi_{\mu}({\bf R}(t_{2}))|\nabla \hat{V}_\mathrm{el} (\hat{\bf r};\mathbf{R}(t_2))|\Phi_{\nu}({\bf R}(t_{2}))\rangle b^{\dagger}_{\beta\nu} \nonumber\\
&=&\sum_{\mu\nu} b_{\alpha\mu} \nabla V_{\mu\nu}({\bf R}(t_2)) b^{\dagger}_{\beta\nu}\\
&~&+\sum_{\nu} E_{\nu} ({\bf R}(t_2))  \langle \Phi_{\alpha}({\bf R_0}) |\nabla\Phi_{\nu}({\bf R}(t_{2}))\rangle b^{\dagger}_{\beta\nu} \nonumber\\
&~&+\sum_{\mu} E_{\mu} ({\bf R}(t_2)) b_{\alpha\mu}  \langle \nabla \Phi_{\mu}({\bf R}(t_{2}))| \Phi_{\beta}({\bf R_0})\rangle\nonumber.
\end{eqnarray}
With $V_{\mu\nu}(\mathbf{R}(t_2))=E_{\mu}({\bf R}(t_2)) \delta_{\mu\nu}$, we realize that Eqn.~\ref{eqn:adgrad} is nothing more than directly applying $\nabla$ on the energy expression $V_{\alpha\beta}({\bf R}(t_{2}))$ in Eqn.~\ref{eqn:elect2}, resulting three terms based on the chain rule. The first term on the right hand side of Eqn.~\ref{eqn:adgrad} is the result of the nuclear dependence on the adiabatic energy, and the last two terms on the right hand side of Eqn.~\ref{eqn:adgrad} are the results of the nuclear dependence on adiabatic orbitals (adiabatic states), weighted by the corresponding adiabatic energies. We emphasize that Eqn.~\ref{eqn:nucgrad} is an equivalent but more compact expression compared to Eqn.~\ref{eqn:adgrad}, which naturally indeed includes derivatives with respect to all possible sources of the nuclear dependence. Further, we emphasize that in the QD propagation scheme, the derivative couplings ${\bf d}_{\mu\nu}({\bf R})=\langle \Phi_{\mu}({\bf R})|\nabla \Phi_{\nu}({\bf R})\rangle$ are not explicitly required. That being said, we do not omit the derivative coupling; the gradient $\langle \Phi_{\mu}({\bf R}(t_{2}))|\nabla \hat{V}_\mathrm{el} (\hat{\bf r};{\bf R}(t_2))|\Phi_{\nu}({\bf R}(t_{2}))\rangle$ used in the QD scheme (Eqn.~\ref{eqn:nucgrad}) is reminiscent of the derivative coupling. One should note that ${\bf d}_{\mu\nu}({\bf R})=\langle \Phi_{\mu}({\bf R})|\nabla \hat{V}_\mathrm{el} (\hat{\bf r};{\bf R})|\Phi_{\nu}({\bf R})\rangle/[E_{\nu}({\bf R})-E_{\mu}({\bf R})]$ can become singular due to the degeneracy of eigenvalues, {\it i.e.}, $E_{\nu}({\bf R})-E_{\mu}({\bf R})=0$, even when $\langle \Phi_{k}({\bf r};{\bf R})|\nabla \hat{V}_\mathrm{el} (\hat{\bf r};{\bf R})|\Phi_{l}({\bf r};{\bf R})\rangle$ is finite. Thus, the method that directly requires derivative couplings might suffer from numerical instabilities, whereas the method only requires the gradient (such as the QD scheme) will likely not.

\begin{figure}
 \centering
  \begin{minipage}[h]{\linewidth}
     \centering
     \includegraphics[width=\linewidth]{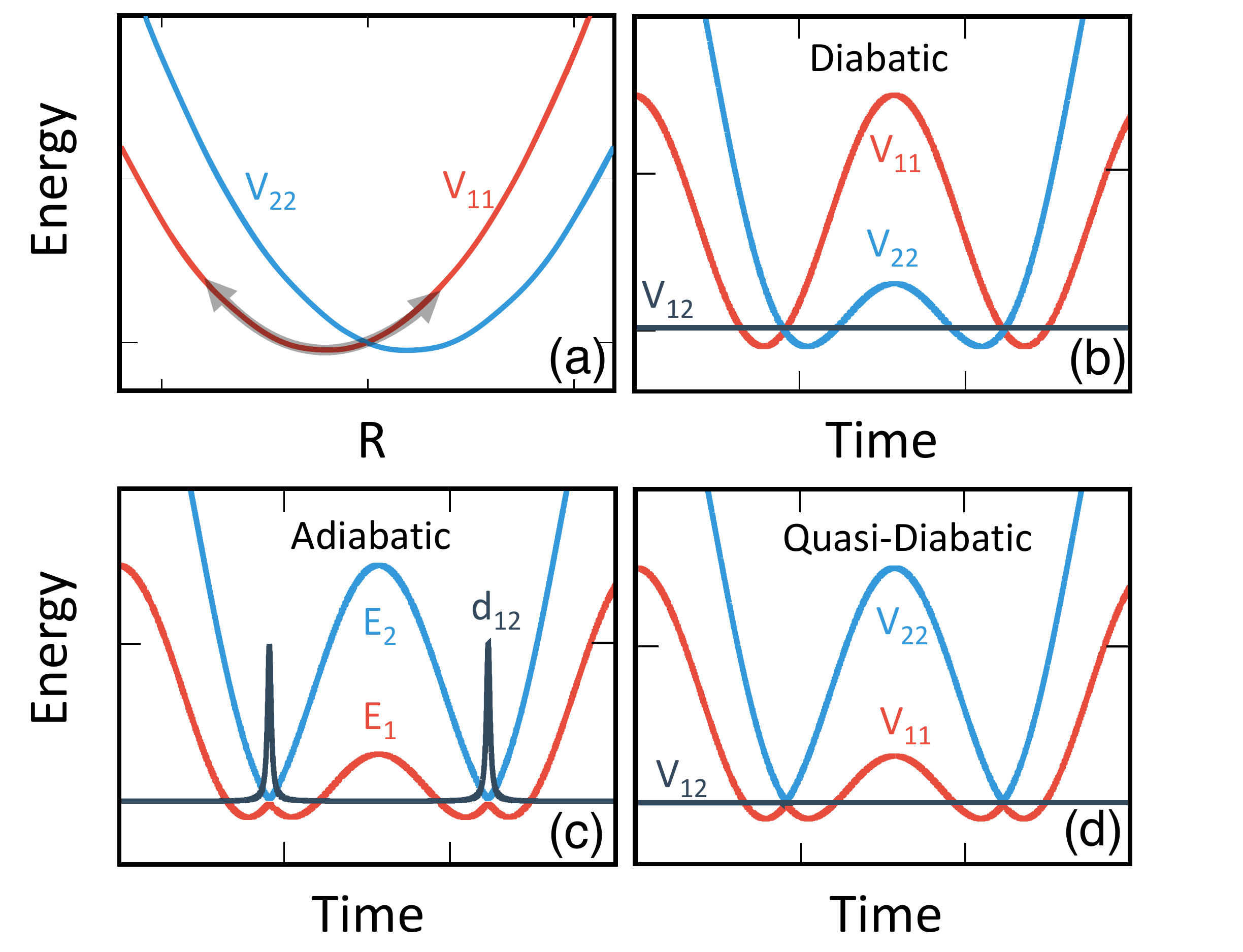}
  \end{minipage}%
   \caption{(a) Diabatic potentials and the corresponding time-dependent potentials and couplings for a one-dimensional model system, in (b) diabatic representation, (c) adiabatic representation, and (d) quasi-diabatic representation.} 
\label{fig:qd}
\end{figure}

Fig.~\ref{fig:qd} presents a simple two-level model system in panel (a) and its time-dependent electronic potential in the (b) diabatic, (c) adiabatic,  and (d) quasi-diabatic representation. In Fig.~\ref{fig:qd}a, the motion of the nuclear trajectory (indicated by the gray double-sided arrow) is confined on the diabatic state 1 (red surface). The diabatic potential energy surfaces $V_{11}(R(t))$ and $V_{22}(R(t))$ presented in panel (b) evolve smoothly in time, and the diabatic electronic coupling $V_{12}$ is a small constant in this model. In panel (c), under the adiabatic representation, the derivative coupling vector $d_{12}(R)=\langle \Phi_{1}(R)|\nabla \Phi_{2}(R)\rangle$ starts to exhibit large peaks at the avoided crossing regions, where the adiabatic wavefunctions rapidly change their characters. These rapid changes of derivative couplings usually cause numerical challenges and require a very small time step for a stable quantum dynamics propagation. The QD representation presented in panel (d), on the other hand, vanishes the derivative couplings; the non-adiabatic transitions are induced by the overlap between two consecutive QD bases $\langle \Phi_{1}(R(t_2))| \Phi_{2}(R(t_1))\rangle$. The off-diagonal electronic coupling $V_{12}(R(t))$ under QD has small values due to the varying QD basis along the propagation (see Eqn.~\ref{eqn:interpolation}), and will decrease to zero under the limit that $(t_2-t_1) \rightarrow 0$. 

Thus, the QD representation provides several unique advantages over the strict diabatic or adiabatic representation for quantum dynamics propagations. On one hand, the QD basis are constructed from the crude adiabatic basis, which can be easily obtained from any commonly used electronic structure calculations. On the other hand, the diabatic nature of the QD basis makes derivative couplings explicitly vanish and allows using any diabatic dynamics approaches to perform on-the-fly propagation. Further, the QD scheme ensures a stable propagation of the quantum dynamics compared to directly solving it in the adiabatic representation. This is due to the fact that directly solving electronic dynamics in the adiabatic state requires the non-adiabatic coupling $\langle \Phi_\beta({\bf R}(t))|{\partial\over{\partial t}}\Phi_\alpha({\bf R}(t))\rangle={\bf d}_{\beta\alpha}({\bf R})\dot{\bf R}$, which might exhibit highly peaked values and cause large numerical errors\cite{Meek2014, Subotnik2016} when using the linear interpolation scheme.\cite{tully94jcp} The QD scheme explicitly alleviates this difficulty by using the well behaved transformation matrix elements $\langle \Phi_\beta({\bf R}(t_1))|\Phi_\alpha({\bf R}(t_2))\rangle$ instead of $\langle \Phi_\beta({\bf R}(t))|{\partial\over{\partial t}}\Phi_\alpha({\bf R}(t))\rangle$. 

Further, we note that the QD scheme was historically introduced for propagating the electronic amplitudes in surface hopping calculations.\cite{Rossky-Webster,yu2001influence,GranucciLD1,GranucciLD2} It has also been used in scattering probability calculations\cite{truhlar:1981} and recently, gaussian wave packet dynamics approaches,\cite{Meek2016JCP,Meek2016JCP2,DmitryJCP2014,FernandezPCCP2015,Izmaylov2018JCP} and is referred as the moving crude adiabatic scheme.\cite{Izmaylov2018JCP} Here, we significantly expand the scope of this scheme\cite{Huo2018} by using it as a general framework to interface any {\it diabatic} trajectory-based dynamics methods with any {\it adiabatic} electronic structure calculations.

Finally, we note that in a real molecular system, the QD state at the nuclear position ${\bf R}(t_2)$ may no longer be a complete basis set. As a consequence, the total population will decay from unity during the dynamical propagation after applying many of the basis transformations. This problem, however, can be easily addressed by performing orthonormalization procedure\cite{Huo2018} among vectors $\{\langle {\bf \Phi}({\bf R}(t_1))| \Phi_{\mu}({\bf R}(t_2))\rangle \}$. The details of this procedure, as well as one specific example can be found in a charge transfer QD dynamics simulation in our original QD work.\cite{Huo2018} This procedure has not been applied to any of the model calculations in this paper.

\subsection{Algorithm for QD-SQC propagation}\label{sec:algorithm}
Combining the diabatic SQC approach and the QD propagation scheme described above, we formulate the following algorithm for the QD-SQC quantum dynamics propagation: 

{\bf 1}. sample the initial conditions of the nuclear DOF ${\bf R}(t_1)$ and ${\bf P}(t_{1}+{\Delta t\over2})$ based on the Wigner distribution $W_{\bf R}({\bf R},{\bf P})$; uniformly sample the mapping action based on the window function $W_{j}({\bf n})$, and the mapping angle variables $\theta_j\in[0,2\pi]$ for all electronic states $|j\rangle$.

{\bf 2}. perform electronic structure calculations at $t_1$ to obtain the QD basis $|\Phi_{\alpha}({\bf R_{0}})\rangle\equiv|\Phi_{\alpha}({\bf R}(t_1))\rangle$.

{\bf 3}. propagate nuclear positions as ${\bf R}(t_{2}) = {\bf R}(t_{1}) + {{\bf P}(t_{1}+{{\Delta t}\over2})}\Delta t/ M$, perform electronic structure calculations at ${\bf R}(t_2)$ to obtain the adiabatic basis $\{|\Phi_{\mu}({\bf R}(t_2))\rangle\}$.

{\bf 4}. compute the electronic Hamiltonian elements $V_{\alpha\beta}({\bf R}(t))$ based on Eqn.~\ref{eqn:interpolation} for $t \in[t_{1}, t_{2}]$, as well as the nuclear gradients $\nabla V_{\alpha\beta}({\bf R}(t_{2}))$ based on Eqn.~\ref{eqn:nucgrad}.

{\bf 5}. propagate the canonical mapping variables $\{{\bf q},{\bf p}\}$ by solving Eqn.~\ref{eq:mapeqn} with the electronic elements $V_{\alpha\beta}({\bf R}(t))$ computed from step {\bf 4}; propagate the nuclear momenta as ${\bf P}\left(t_{2}+\frac{\Delta t}{2}\right) = {\bf P}(t_{1}+\frac{\Delta t}{2}) + {\bf F}({\bf R} (t_{2})) \Delta t/M$, with the force computed at nuclear position ${\bf R} (t_{2})$ based on Eqn.~\ref{eqn:nucforce}.

{\bf 6}. transform the canonical mapping variables from the instantaneous QD basis $\{q_\alpha,p_\alpha\}$ back to the strict diabatic basis $\{q_i,p_i\}$, with $q_{i} =\sum_{\alpha}\langle  \Phi_{\alpha}({\bf R_0})|i \rangle q_\alpha$ and $p_{i} =\sum_{\alpha}\langle \Phi_{\alpha}({\bf R_0})|i \rangle p_\alpha$; compute the action variables based on $n_i = \frac{1}{2}\left(p_i^2 + q_i^2 - 2\gamma \right)$; evaluate the strict diabatic populations with the Window function estimator in Eqn.~\ref{eqn:square} or Eqn.~\ref{eqn:triangle}, and renormalize population based on Eqn.~\ref{eq:popNormalize}.

{\bf 7}. transform the mapping variables into the new QD basis $|\Phi'_{\mu}({\bf R'_{0}})\rangle\equiv|\Phi_{\mu}({\bf R}(t_2))\rangle$ for the $[t_2,t_3]$ propagation step, with the following expressions: $\sum_{\alpha} q_{\alpha} \langle \Phi_{\alpha}({\bf R}(t_{1}))| \Phi_{\mu}({\bf R}(t_{2}))\rangle  \rightarrow q_{\mu}$ and $\sum_{\alpha} p_{\alpha} \langle \Phi_{\alpha}({\bf R}(t_{1}))| \Phi_{\mu}({\bf R}(t_{2}))\rangle  \rightarrow p_{\mu}$.

{\bf 8}. repeat steps {\bf 3}-{\bf 7}.
\\

Here, we want to comment on three technical details for the QD propagation scheme. First, we have transformed the mapping variables between two bases in step {\bf 6} and {\bf 7}. This process is valid because the relation between two QD bases in step 7 are $|\Phi_{\mu}({\bf R}(t_{2}))\rangle=\sum_{\alpha} \langle \Phi_{\alpha}({\bf R}(t_{1}))| \Phi_{\mu}({\bf R}(t_{2}))\rangle|\Phi_{\alpha}({\bf R}(t_{1}))\rangle$. Since the mapping relation between the physical state and the singly excited oscillator state is $|\Phi_{\mu}({\bf R}(t_{2}))\rangle=a_{\mu}^{\dagger}|0\rangle=\frac{1}{\sqrt{2}}(\hat{q}_\mu + i\hat{p}_\mu)|0\rangle$, the relations for the mapping variables associated with two bases are $|\Phi_{\mu}({\bf R}(t_{2}))\rangle=\frac{1}{\sqrt{2}}(\hat{q}_\mu + i\hat{p}_\mu)|0\rangle=\sum_{\alpha} \langle \Phi_{\alpha}({\bf R}(t_{1}))| \Phi_{\mu}({\bf R}(t_{2}))\rangle\frac{1}{\sqrt{2}} (\hat{q}_\alpha + i\hat{p}_\alpha)|0\rangle$. For molecular systems, one can always find a suitable choice for the basis set in order to make $\langle \Phi_{\alpha}({\bf R}(t_{1}))| \Phi_{\mu}({\bf R}(t_{2}))\rangle$ real, which guarantees that the mapping variables are transformed with the same relations as the bases. Similarly, in step {\bf 6}, we transform the time-dependent mapping variables from the instantaneous QD basis, $\{q_\alpha,p_\alpha\}$, to the strict diabatic basis, $\{q_i,p_i\}$. Note that the relation between the strict diabatic \{$|i\rangle$\} and QD \{$|\Phi_{\alpha}({\bf r};{\bf R_0}) \rangle$\} states are $|i \rangle =\sum_{\alpha}|\Phi_{\alpha}({\bf R_0}) \rangle\langle \Phi_{\alpha}({\bf R_0})|i \rangle$, which leads to the following transformations for  mapping variables associated with two bases
\begin{equation}
q_{i} =\sum_{\alpha}\langle  \Phi_{\alpha}({\bf R_0})|i \rangle q_\alpha;~~~p_{i} =\sum_{\alpha}\langle \Phi_{\alpha}({\bf R_0})|i \rangle p_\alpha.
\end{equation}

Second, the nuclear force evaluated in the QD basis in step {\bf 5} has the same form of the nuclear force in the strict diabatic basis $\{|i\rangle,|j\rangle\}$. This is valid based on the following analysis. Consider expanding the strict diabatic basis as the linear combination of QD basis, with $|i\rangle=\sum_{\alpha}|\Phi_{\alpha}({\bf R_{0}})\rangle\langle\Phi_{\alpha}({\bf R_{0}})|i\rangle=\sum_{\alpha}C_{i\alpha}|\Phi_{\alpha}({\bf R_{0}})\rangle$. This implies that $q_i=\sum_{\alpha}C_{i\alpha}q_\alpha$ and $p_i=\sum_{\alpha}C_{i\alpha}p_\alpha$. Plugging in these two expressions into the nuclear force ${\bf F}=-{1\over 2}\sum\limits_{ij}\nabla V_{ij}(R)[p_{i}p_{j}+q_{i}q_{j}-2\gamma \delta_{ij}]$ in the diabatic representation, and noticing the fact that $\delta_{ij}=\langle i|j\rangle$ as well as explicitly using the transformation relation among states, we obtain the nuclear forces in the QD representation as follows
\begin{eqnarray}\label{eq:force}
{\bf F}&=&-{1\over 2}\sum\limits_{ij\alpha \beta } C_{i\alpha}\nabla V_{ij}(R)C_{j\beta}\nonumber\\
&&~~\times[p_{\alpha}p_{\beta}+q_{\alpha}q_{\beta} -2\gamma\langle \Phi_{\alpha}({\bf R_{0}})|\Phi_{\beta}({\bf R_{0}})\rangle]\nonumber\\
&=&-{1\over 2}\sum\limits_{\alpha\beta} \nabla V_{\alpha\beta}(R) \left[p_{\alpha}p_{\beta}+q_{\alpha}q_{\beta}-2\gamma \delta_{\alpha \beta}\right], \nonumber
\end{eqnarray}
which indeed has the nuclear force expression in the diabatic representation as described in Eqn.~\ref{eqn:nucforce}. Note that in the above equation, we use the fact that $\sum_{ij}C_{i\alpha}\nabla V_{ij}(R)C_{j\beta}=\sum_{ij}C_{i\alpha}\langle i|\nabla \hat{V}(R)|j\rangle C_{j\beta}=\langle\Phi_{\alpha}({\bf R_{0}})|\nabla \hat{V}(R)|\Phi_{\beta}({\bf R_{0}})\rangle$.

Third, in step {\bf 6} we evaluate the population with the window function defined in the strict diabatic basis $\{|i\rangle, |j\rangle\}$, despite that the mapping trajectories are propagated in the quasi-diabatic basis (in QD-SQC). For strict diabaitc model systems, we use the above procedure to demonstrate that when the populations are evaluated with the {\it same} diabatic window functions, the results obtained from QD-SQC propagations are exactly the same as those obtained from the diabatic SQC propagations. In real molecular systems, however, the strict diabatic states are not easily obtained. Thus, one might have to compute the population with the window function defined in the QD states, which are those instantaneous adiabatic states used to define the QD states. In this paper, we test the performance of QD-SQC in this scenario with a photoinduced proton-coupled electron transfer (PI-PCET) model system,\cite{Hazra:2010,Hazra:2011} where the proton-electron adiabatic vibronic states are used as the QD states, and there is no obvious exact low-dimensional diabatic vibronic state without further applying diabatization procedures.\cite{Ananth:2017,Kretchmer:2013} For this particular application, we use the window function defined in the QD basis (or equivalently in the instantaneous adiabatic states). Thus, the shape of the window function will change along the nuclear trajectory ${\bf R}(t)$, instead of a fixed shape when it is defined in the strict diabatic states (Fig.~\ref{fig:sqc}). We will fully explore the consequence of using such adiabatic window functions in future investigations. 

Finally, we would like to emphasize that the accuracy of QD-SQC will be limited by the accuracy of SQC itself, {\it i.e.,} the validity of using a window function as an approximate pre-limit delta function, as well as the Ehrenfest-type mean-field dynamics. The QD propagation scheme is rather general and provides a convenient framework that allows interfacing {\it diabatic} dynamics approaches with {\it adiabatic} electronic structure calculations for on-the-fly propagation.\cite{Huo2018}

\section{Details of Model Calculations}
\subsection{Diabatic Models}
\begin{figure}
 \centering
  \begin{minipage}[t]{\linewidth}
     \centering
     \includegraphics[width=\linewidth]{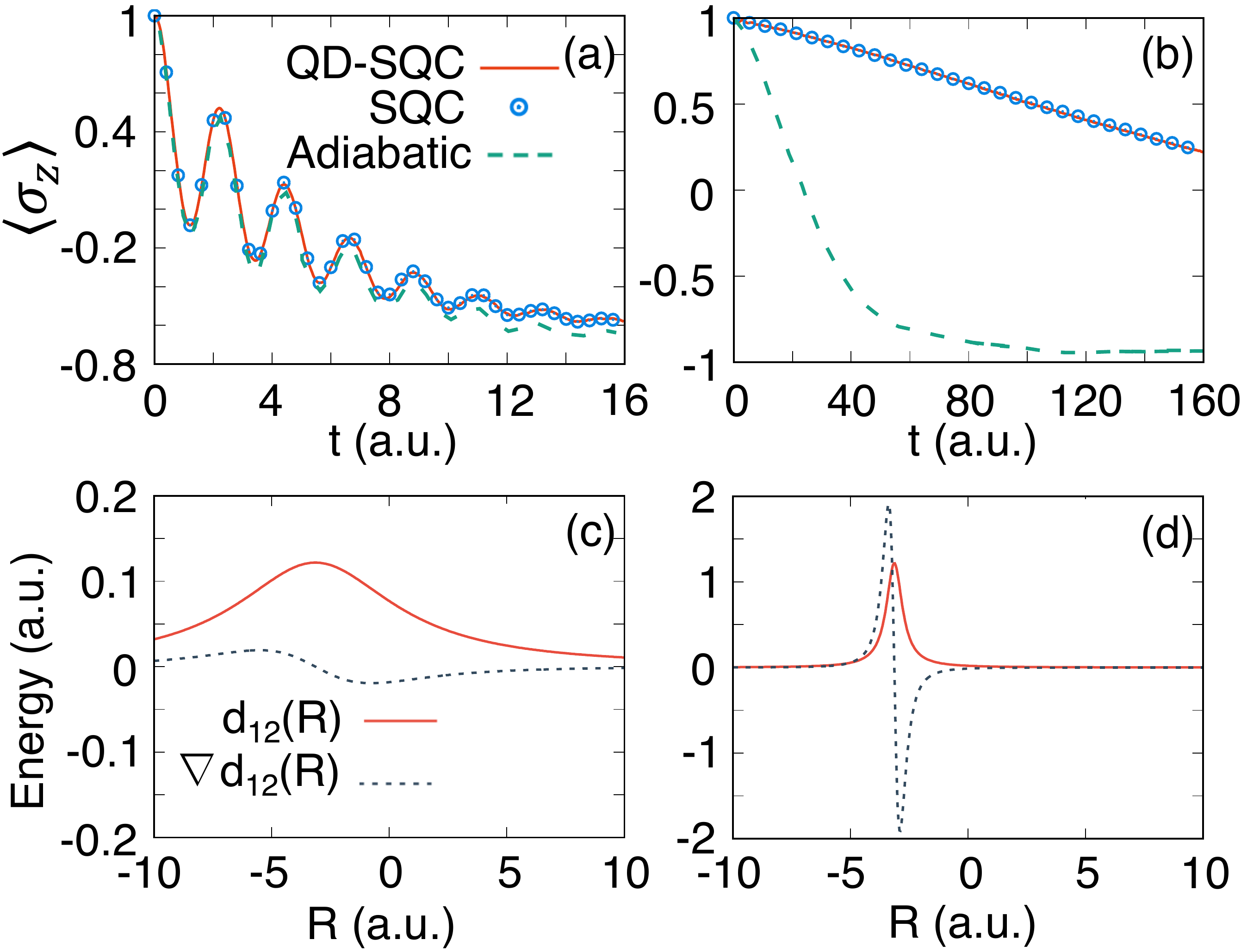}
  \end{minipage}%
   \caption{Population dynamics of the spin-boson model in (a) strong (b) weak diabatic electronic coupling regime, obtained from SQC calculation in the diabatic (open circles), adiabatic (dash), and quasi-diabatic (solid) representation. In the adiabatic SQC calculation, the second derivative couplings are ignored.\cite{MillerAdiabatic} Derivative coupling $d_{12}(R)$ and its derivative $\nabla d_{12}(R)$  are calculated for both scenarios and presented in panels (c) and (d). The $R$ coordinate is chosen based on its frequency, which is closest to $\omega_c$.}
\label{sb-adiabatic}
\end{figure}

Here, we provide details of the calculations with the strict diabatic model systems. These models are carefully chosen to include a wide range of scenarios in non-adiabatic dynamics, such as weak/strong avoid crossings, conical intersection, and many-state system, in order to fully assess the performance of the QD scheme. In this section, we provide details of the spin-boson model (Fig.~\ref{sb-adiabatic} and Fig.~\ref{spin}) and an excitation energy transfer model\cite{ishizaki2008pnas} (Fig.~\ref{fmo}). The conical intersection model\cite{StockConical, StockConicalJCP95}  (Fig.~\ref{conical}) is discussed in the result section, and the details for Tully's avoid crossing Model systems can be directly found in literature.\cite{Tully} 

{\bf Model Hamiltonians}. The spin-boson model has the Hamiltonian $\hat{H}=\sum_{k}[\hat{P}_{k}^2/2+\omega^2_{k}\hat{R}^2/2+c_{k}\hat{R}_{k}\hat{\sigma}_{z}]+ \epsilon\hat{\sigma}_{z}/2+\Delta \hat{\sigma}_{x}$, with electronic bias $\epsilon$, electronic coupling $\Delta$, and the system-bath coupling $c_{k}$ for a given spectral density $J(\omega)={\pi\over2}\sum_{k}{c^2_{k}\over\omega_{k}}\delta(\omega-\omega_k)$. Here, we use 100 discretized harmonic modes to sample\cite{Makri99} the spectral density $J(\omega)={\pi\over2}\xi\omega e^{-\omega/\omega_c}$, where $\xi$ is the Kondo parameter and $\omega_c$ is the cut-off frequency (peak of the spectral density). For the model calculations in this paper, we use $\Delta = 1$ and $\omega_c = 2.5$ (Fig.~\ref{spin}a-c) or $\omega_c = 1$ (Fig.~\ref{spin}d). The initial Wigner distribution for the bath modes is centered around ${R}_{k}(0) =-{c_k}/\omega_{k}^2$ and ${P}_{k}(0) =0$.

For simulating singlet excitation energy transfer in a dissipative environment, we use the following Frenkel exciton model $\hat{H}=\hat{H}_\mathrm{ex}+\hat{H}_\mathrm{sb}$.  The exciton part of the Hamiltonian is $\hat{H}_\mathrm{ex}=\sum_{i}\epsilon_{i} |i \rangle \langle i|+\sum_{i\neq j}\Delta_{ij} |i\rangle \langle j |$, with singlet excitation energy $\epsilon_{i}$ on chromophore $i$ and the electronic coupling $\Delta_{ij}$  between two single excitations $|i\rangle$ and $|j\rangle$. The system-bath Hamiltonian that describes the exciton-phonon interactions is $\hat{H}_\mathrm{sb}=\sum_{i}\sum_{k_{i}}[{1\over 2} (\hat{P}^2_{k_{i}} + \omega^2_{k_{i}}\hat{R}^2_{k_{i}}) + c_{k_{i}}\hat{R}_{k_{i}} | i \rangle\langle i |]$, where each state $|i\rangle$ is coupled to a set of independent harmonic bath modes $\{R_{k_{i}}\}$. Here, we use the model parameters of the Fenna-Matthews-Olson (FMO) complex that contains seven chromophores.\cite{ishizaki2008pnas} In addition, we use 60 modes to sample the spectral density $J(\omega)= 2\lambda\omega\tau/(1+(\omega\tau)^2)$ for each bath, where the reorganization energy is $\lambda=35$ cm$^{-1}$, and the solvent response time is $\tau=$50 fs. The parameters for $\hat{H}_\mathrm{ex}$ can be found in Ref. \citen{ishizaki2008pnas} and the sampling procedure for the spectral density can be found in Ref. \citen{MillerJCTC16}. The initial Wigner distribution for each bath mode is centered around ${\bf R}(0)=0$ and ${\bf P}(0)=0$.

{\bf Electronic Matrix Elements for the QD propagation.} For these diabatic model systems, the matrix elements of the electronic Hamiltonian $V_{ij}({\bf R}(t))$ and the nuclear gradients in the diabatic representation $\nabla V_{ij}({\bf R}(t))$ are available and directly used in SQC propagations. For QD-SQC propagations, the adiabatic basis $\{|\Phi_{\alpha}({\bf r};{\bf R}(t))\rangle\}$ is obtained by diagonalizing $V_{ij}({\bf R}(t))$ matrix, which is used as the QD basis. The matrix elements of the electronic Hamiltonian and nuclear gradients are evaluated using Eqn.~\ref{eqn:vijt}-\ref{eqn:elect2} and Eqn.~\ref{eqn:nucgrad}, respectively. Alternatively, these elements can be easily computed by taking advantage of the available diabatic basis in all of our model calculations, for example, as $V_{\alpha\beta}({\bf R}(t))  = \sum_{ij}\langle \Phi_\alpha ({\bf R_0})|i\rangle V_{ij}({\bf R}(t))\langle j|\Phi_\beta({\bf R_0})\rangle$. Both protocols generate the same results.

{\bf Initial Conditions}. The initial conditions for all of the model calculations are $\hat{\rho}(0)=|i\rangle \langle i|\otimes \hat{\rho}_{\bf R}$, where $|i\rangle$ indicates the initial electronic diabatic state and $\hat{\rho}_{\bf R}$ represents the initial nuclear density operator. For non-adiabatic scattering and photo-dissociation calculations presented in Fig.~\ref{tully}-\ref{conical}, we use $\hat{\rho}_{\bf R}=|\chi\rangle\langle \chi|$, where $\langle {\bf R}|\chi \rangle=\left({{2\Gamma}\over\pi}\right)^{1/4}e^{-(\Gamma/2)({\bf R}-{\bf R}_{0})^2+{i\over\hbar}{\bf P}_{0}({\bf R}-{\bf R}_{0})}$ represents a Gaussian wavepacket centered around ${\bf R_0}$ and ${\bf P_0}$ with a width $\Gamma$. The corresponding nuclear Wigner density is $W_{\bf R}({\bf P},{\bf R})={1\over{\pi}}e^{-\Gamma({\bf R}-{\bf R_{0}})^2 -({{\bf P}-{\bf P_0})^2/\Gamma}}$. For the condensed-phase model calculations presented in Fig.~\ref{fmo}-\ref{spin}, we assume that each nuclear DOF is represented by a harmonic mode. The canonical thermal density for the $k_\mathrm{th}$ nuclear DOF $R_{k}$ is thus $\hat{\rho}_{\bf R}(\hat{P}_{k},\hat{R}_{k})={1\over \mathcal{Z}}e^{-{1\over{k_\mathrm{B}T}}[\hat{P}_{k}^{2}/2M+{1\over 2}M\omega_{k}^{2}\hat {R}_{k}^{2})]}$. The corresponding nuclear Wigner density is then $W_{\bf R}({P}_k,{R}_k)=2\tanh({{\omega_k}\over{2\mathrm{k_{B}}T}})e^{-\tanh({{\omega_k}\over{2\mathrm{k_{B}}T}})\left[m\omega_{k}(R_{k}-R_{k}(0))^2+{P}_{k}(0)^2/(m\omega_k)\right]}$. 

{\bf Window Functions and Convergence}. In this paper, all of the calculations for two level systems are performed with the triangle window function (Eqn.~\ref{eqn:triangle}), which has been shown to provide the accurate electronic dynamics across a broad range of the electronic couplings.\cite{CottonJCP16} The only results obtained with square window function are the seven-states excitation energy transfer calculations\cite{ishizaki2008pnas} presented in Fig.~\ref{fmo}, and the adiabatic vibronic dynamics of the model PI-PCET system presented in Fig.~\ref{fig:pipcet}.  All of the results for diabatic model Hamiltonian are obtained with 24,000 trajectories, except those in FMO model (Fig.~\ref{fmo}) where 200,000 trajectories are used. The same time step $dt$ are used for both SQC and QD-SQC calculations.

\subsection{Adiabatic Vibronic Model}
Here, we provide details of the adiabatic vibronic dynamics calculations with a PI-PCET model system,\cite{Hazra:2010,Hazra:2011} presented in Fig.~\ref{fig:pipcet}. More details about this model can also be found in our recent work.\cite{mandal2018} Despite its simple form (which contains strict diabatic electronic states), this model provides a more stringent test of the QD-SQC approach because (without further diabatization procedure) there is no obvious low-dimensional diabatic vibronic states.

{\bf Model Hamiltonian.} The PI-PCET model used in this study is expressed as $\hat{H}=\hat{H}_\mathrm{ep}+\hat{H}_\mathrm{sb}$, where $\hat{H}_\mathrm{ep}$ describes the electron-proton free-energy surfaces, and $\hat{H}_\mathrm{sb}$ describes the solvent-bath interaction. In this paper, we focus on a symmetric PI-PCET system with zero driving force (bias) of the reaction. The electron-proton Hamiltonian $\hat{H}_\mathrm{ep}$ defined in the electronic diabatic representation $\{|\mathrm{D}\rangle,|\mathrm{A}\rangle\}$ is expressed as 
\begin{equation}\label{eqn:langeham}
\resizebox{1.0\hsize}{!}{
$\hat{H}_\mathrm{ep}=\hat{T}_\mathrm{p}+$
\Bigg[
  \begin{tabular}{cc}
  $U^\mathrm{D}(\hat{r}_\mathrm{p})+\frac{1}{2}M_\mathrm{s}\omega^{2}_\mathrm{s} R_\mathrm{s}^2$ & $V_\mathrm{DA}$\\
   $V_\mathrm{DA}$ & $U^\mathrm{A}(\hat{r}_\mathrm{p})+\frac{1}{2}M_\mathrm{s}\omega^{2}_\mathrm{s} (R_\mathrm{s}- R_\mathrm{s}^0)^2$\\
  \end{tabular}
  \Bigg].}
\end{equation}
Here, $\hat{T}_\mathrm{p}$ represents the kinetic energy operator of the proton, $\hat{r}_\mathrm{p}$ is the proton coordinate operator, and $R_\mathrm{s}$ represents the collective solvent coordinate that characterizes electron transfer. In addition, $M_\mathrm{s}$ and $\omega_\mathrm{s}=\sqrt{f_0/M_\mathrm{s}}$ are the mass and the frequency of this solvent coordinate, with $f_0$ as the force constant, and $R^0_\mathrm{s} =\sqrt{2\lambda/f_0}$, with $\lambda$ as the solvent reorganization energy. The second term of Eqn.~\ref{eqn:langeham}, {\it i.e.}, $\hat{H}_\mathrm{ep}-\hat{T}_\mathrm{p}$ operator, represents the electron-proton interaction potential in the electronic diabatic donor $|\mathrm{D} \rangle$ and acceptor $|\mathrm{A} \rangle$ excited states, with $V_\mathrm{DA}=0.03$ eV as the coupling between the two electronic states. The excited adiabatic states $|\mathrm{S_1}(R_\mathrm{s}, r_\mathrm{p}) \rangle$ and $|\mathrm{S_2} (R_\mathrm{s}, r_\mathrm{p})\rangle$ are the eigenstates of the $\hat{H}_\mathrm{ep}-\hat{T}_\mathrm{p}$ operator, {\it i.e.}, they are linear combinations of $|\mathrm{D} \rangle$ and $|\mathrm{A} \rangle$ states, and are parametrically dependent on both the solvent and the proton coordinates. The electronic ground state $|\mathrm{S_0} (R_\mathrm{s}, r_\mathrm{p}) \rangle$ of the system, on the other hand, is not explicitly included in this  Hamiltonian, but it will dictate the initial conditions of the system before the Franck-Condon photoexcitation.

Further, $U^\mathrm{D}(\hat{r}_\mathrm{p})=\frac{1}{2}m_{p}\omega_\mathrm{p}^{2} (\hat{r}_\mathrm{p}-r_\mathrm{p}^\mathrm{D})^{2}$ and $U^\mathrm{A}(\hat{r}_\mathrm{p})=\frac{1}{2}m_{p}\omega_\mathrm{p}^{2} (\hat{r}_\mathrm{p}-r_\mathrm{p}^\mathrm{A})^{2}$ represent the proton free-energy profile associated with $|\mathrm{D}\rangle$ and $|\mathrm{A}\rangle$ states. In this work, we use $r_\mathrm{p}^\mathrm{D}=0$ and $r_\mathrm{p}^\mathrm{A}=0.5$ \AA~as the minima of proton free-energy profile associated with the electronic donor and acceptor states. $m_\mathrm{p}=1.0073$ amu and $\omega_\mathrm{p}=3000~\mathrm{cm}^{-1}$ are the mass and vibrational frequency of the proton. In this model, the proton and the solvent DOF do not explicitly interact with each other; rather, $\hat{r}_\mathrm{p}$ directly interacts with various electronic states, which in turn interact with the solvent. All the other parameters are provided in Appendix B.

The solvent-bath Hamiltonian $\hat{H}_\mathrm{sb}$ is expressed as follows
\begin{equation}\label{eqn:system-bath}
\hat{H}_\mathrm{sb}=\frac{P^{2}_\mathrm{s}}{2M_\mathrm{s}}+\sum_{k}\left[\frac{P_{k}^2}{2M_{k}} + \frac{1}{2}M_{k} \omega_{k}^{2} \left(R_{k}- \frac{c_{k}R_\mathrm{s}}{M_{k} \omega_{k}^{2}}\right)^2\right].
\end{equation}
In the above equation, $R_k$ represents the $k_\mathrm{th}$ bath mode, with the corresponding coupling constant $c_k$ and frequency $\omega_k$ sampled from the following spectral density $J(\omega)={\pi\over2}\sum_{k}{{c^{2}_{k}}\over{M_{k}\omega_k}}\delta(\omega-\omega_k)=f_0\tau_{\text{L}}\omega e^{-{{\omega}\over{\omega_\mathrm{c}}}}$. Here, $\tau_{\text{L}}$ is the solvent response time (see Appendix B), $M_{k}$ is the mass of the $k_\mathrm{th}$ bath mode, and $\omega_\mathrm{c}$ is the characteristic frequency of the bath that is much faster than the motion of $R_\mathrm{s}$. Here, we choose $\omega_\mathrm{c}=10\omega_\mathrm{s}$ and $M_{k}=M_\mathrm{s}$ for all $k$. One can thus perform QD-SQC simulation with the above total Hamiltonian. 

Instead of treating the bath DOF explicitly, we can perform the following equivalent Langevin dynamics\cite{Hazra:2010} to implicitly treats the influence of the bath, with the equation of motion for the collective solvent coordinate $R_\mathrm{s}$ as follows
\begin{equation}\label{eqn:langevin}
M_\mathrm{s} \ddot{R_\mathrm{s}}={\bf F}_\mathrm{ep}(R_\mathrm{s})-f_0 \tau_{\text{L}}\dot{R_\mathrm{s}}+{\bf F}_\mathrm{r}(t). 
\end{equation}
In the above Langevin equation, ${\bf F}_\mathrm{ep}(R_\mathrm{s})$ is the SQC nuclear force (see Eqn.~\ref{eqn:nucforce}) evaluated with $\hat{H}_\mathrm{ep}$, the friction force is $-f_0 \tau_{\text{L}}\dot{R_\mathrm{s}}$ with the friction constant $f_0 \tau_{\text{L}}$, and ${\bf F}_\mathrm{r}(t)$ is the random force bounded by the fluctuation-dissipation theorem through equation $\langle {\bf F}_\mathrm{r}(t){\bf F}_\mathrm{r}(0)\rangle=2k_\mathrm{B}Tf_0 \tau_{\text{L}}\delta(t)$.  Here, $\textbf{F}_{\text{r}}(t)$ is modeled as a Gaussian random force with the distribution width\cite{Tully:1979} $\sigma =\sqrt{2k_{\text{B}}T f_0 \tau_{\text{L}}/dt}$, where $k_{\text{B}}$ is the Boltzmann constant and $dt$ is the nuclear time step. The details for generating $\tau_\mathrm{L}$ for a given solvent is provided in Appendix B. As a consistency check, in our previous work of using QD partial-linearized density matrix (QD-PLDM) approach for simulating this model,\cite{mandal2018} we have verified that equivalent results (for time-dependent electron-proton reduced density matrix) are obtained with either explicit bath (dynamics with the full Hamiltonian $\hat{H}_\mathrm{ep}+\hat{H}_\mathrm{sb}$) or implicit bath (Langevin dynamics in Eqn.~\ref{eqn:langevin}) approach. The equivalency of both approaches have also been recently explored in the condensed-phase ET dynamics.\cite{HuoJCP13, Subotnik:2013}

{\bf Adiabatic Vibronic Surfaces for QD-SQC propagation.} Here, we treat both electron and proton quantum mechanically with their corresponding vibronic states. The ``electronic Hamiltonian'' $\hat{V}$ in Eqn.~\ref{eq:eigenvalue} is then defined as $\hat{H}_\mathrm{ep}$ in Eqn.~\ref{eqn:langeham}, such that $\hat{V}\equiv \hat{H}_\mathrm{ep}(\hat{T}_\mathrm{p},\hat{r}_\mathrm{p}, \hat{r}_\mathrm{e},R_\mathrm{s})$. Thus, the ``electronic Hamiltonian" $\hat{V}$ includes proton kinetic energy, electronic potential, as well as electron-proton and electron-solvent interactions, except the nuclear kinetic energy of the solvent (which is treated as the classical DOF).

In order to obtain the adiabatic vibronic states $|\Phi_{\alpha}({R}_\mathrm{s})\rangle$ for the coupled electron-proton Hamiltonian $\hat{H}_\mathrm{ep}$, we express $|\Phi_{\alpha}({R}_\mathrm{s})\rangle$ with a set of two-particle basis functions as follows
\begin{equation}\label{wfexp}
|\Phi_{\alpha}({R}_\mathrm{s})\rangle=\sum\limits_{i,m} c_{im}^{\alpha}({R}_\mathrm{s})|\phi_\mathrm{e}^i\rangle|\phi_\mathrm{p}^m\rangle, 
\end{equation}
where  $|\phi_\mathrm{e}^i\rangle \in \{|\mathrm{D}\rangle,|\mathrm{A}\rangle\}$ and $|\phi_\mathrm{p}^{m}\rangle$ is chosen to be the $m_\mathrm{th}$ eigenfunction of a quantum harmonic oscillator, with the total Hamiltonian $\hat{H}=\hat{T}_\mathrm{p}+{1\over 2}m_\mathrm{p}\omega_ \mathrm{p}^{2}\hat{r}_\mathrm{p}^2$. Thus, by using $M$ harmonic basis functions for proton and two basis states for electron, the total number of vibronic basis is $N=2M$, and $\hat{H}_\mathrm{ep}$ contains $2M\times 2M$ Hamiltonian matrix $\langle\phi_\mathrm{p}^n|\langle\phi_\mathrm{e}^j|\hat{H}_\mathrm{ep}|\phi_\mathrm{e}^i\rangle|\phi_\mathrm{p}^m\rangle$ under this representation. In the model calculation presented in this study, the total number of vibrational basis $\{|\phi_\mathrm{p}^{m}\rangle\}$ is 30, {\it i.e.}, $m=0,1,...29$ for the Harmonic oscillator eigenstates.  

Because both $U^\mathrm{D}(\hat{r}_\mathrm{p})$ and $U^\mathrm{A}(\hat{r}_\mathrm{p})$ are just simple displaced harmonic oscillator potentials, the matrix elements of $\hat{H}_\mathrm{ep}$ can be obtained analytically by recognizing the basic property of harmonic oscillator as follows 
\begin{eqnarray}
&&\langle \phi_\mathrm{p}^n|\hat{T}_\mathrm{p}+{1\over 2}m_\mathrm{p}\omega_\mathrm{p}^{2}\hat{r}_\mathrm{p}^2|\phi_\mathrm{p}^m\rangle=\bigg(n+{1\over2}\bigg)\hbar\omega_\mathrm{p}\delta_{nm} \nonumber\\
&&\langle \phi_\mathrm{p}^n|\hat{r}_\mathrm{p}|\phi_\mathrm{p}^m\rangle = \sqrt{\frac{\hbar}{m_\mathrm{p}\omega_\mathrm{p}}}\frac{1}{\sqrt{2}} \bigg(\sqrt{m}\, \delta_{n,m-1}+\sqrt{m+1}\, \delta_{n,m+1}\bigg).\nonumber
\end{eqnarray}
The eigenvalues and the eigenvectors ({\it adiabatic vibronic} basis) are then obtained through direct diagonalization of the $\hat{H}_\mathrm{ep}$ matrix under the above two-particle basis. 

{\bf Initial Conditions}. The system is initially prepared in the proton vibrational ground state $|\phi_\mathrm{p}^{0}\rangle$ of the electronic ground state $|\mathrm{S_0}\rangle$. The system is then excited to the $|\mathrm{D}\rangle$ state (which is an electronic excited state) through Franck-Condon process, which generate the initial state described by the total density operator $\hat{\rho}(0)=|\Phi(0)\rangle \langle\Phi(0)|\otimes \hat{\rho}_\mathrm{s}$. Here, the initial electron-proton quantum state is expressed as $|\Phi(0)\rangle=|\mathrm{D}\rangle|\phi_\mathrm{p}^{0}\rangle$, and $\hat{\rho}_\mathrm{s}$ is the density operator of the solvent.

In order to initialize the SQC calculation, we need to give a initially occupied state, as the current SQC approach is developed to handle such initial condition.\cite{MillerJCP13} In the model system we studied here, we find that $|\Phi(0)\rangle$ are always nearly identical to one specific adiabatic vibronic state, $|\Phi_{\alpha}({R}_\mathrm{s})\rangle$, such that $\langle\Phi(0)|\Phi_{\beta}({R}_\mathrm{s})\rangle=\delta_{\alpha\beta}$. Note that with different solvent coordinate ${R}_\mathrm{s}$, the corresponding initially occupied adiabatic vibronic states $|\Phi_{\alpha}({R}_\mathrm{s})\rangle$ are different. Nevertheless, it allows using the normal SQC procedure to sample the initial action variables with the square window function $W_{\alpha}$ (Eqn.~\ref{eqn:square}) that corresponds to initially occupied adiabatic vibronic state $|\Phi_{\alpha}({R}_\mathrm{s})\rangle$ for every single trajectory.

The initial configuration of the solvent coordinate is sampled through Wigner density $\rho_\mathrm{s}^\mathrm{W} = { \omega_\mathrm{s} \Gamma_\mathrm{s}} e^{-\Gamma_\mathrm{s} \big[{{P_\mathrm{s}^2\over {2M_\mathrm{s}}} + {1\over2}M_\mathrm{s}\omega_\mathrm{s}^2({R_\mathrm{s}}-{R_\mathrm{s}^0})^2\big]}}$. Here,  $\Gamma_\mathrm{s}= (2/\omega_\mathrm{s})\tanh(\omega_\mathrm{s}/2k_\mathrm{B}T)$ and $\omega_\mathrm{s} = \sqrt{f_0/M_\mathrm{s}}$. In this study, we choose $R^0_\mathrm{s} =\sqrt{2\lambda/f_0}$ that corresponds to the minimum of the proton acceptor free energy diabatic surface. 

{\bf Window Functions and Convergence.} For the PI-PCET model calculation, we use the square window scheme. The converged QD-SQC results of the vibronic dynamics are obtained with 4000 trajectories and a time step of $dt=0.024$ fs (1 a.u.). The trend of the population dynamics, on the other hand, can be obtained with just a few hundred trajectories for this model, comparable to the numerical cost of the widely used fewest-switches surface hopping (FSSH) approach.\cite{Hazra:2010} Instead of mapping a large number of diabatic vibrational basis $\{|\phi_\mathrm{p}^{m}\rangle\}$ (with 30 total vibrational basis in Eqn.~\ref{wfexp}) as been done in recent SQC\cite{Subotnik2018-VER,coker2018} or extended SQC\cite{Geva2018} studies, here, we map the adiabatic electron-proton vibronic states $\{|\Phi_{\alpha}({R}_\mathrm{s})\rangle\}$ with the MMST mapping variables through Eqn.~\ref{eq:mmst}. With the QD-SQC approach developed in this work, we can directly use the diabatic SQC approach\cite{MillerJCP13,SCFaradayMiller} to propagate dynamics with quantities evaluated in the adiabatic vibronic state through the QD scheme.\cite{Huo2018} Further, we are aware that by including more vibronic states, SQC might encounter intrinsic difficulties to fully converge, as been demonstrated by a recent study of vibrational relaxation process in a simple harmonic oscillator.\cite{Subotnik2018-VER} To avoid this potential issue, here we only included the lowest four adiabatic vibronic states, as the photoinduced vibrational relaxation dynamics mainly occurs within these states in this model.\cite{Hazra:2010,mandal2018} To compute the adiabatic vibronic population, we use the window function defined in the instantaneous adiabatic vibronic states (which are also the QD states) to bin the mapping action variables, as oppose the step {\bf 6} in the QD-SQC propagation algorithm in Section.~\ref{sec:algorithm} when well-defined diabatic states exist.

\section{Results and Discussions}
Fig.~\ref{sb-adiabatic} presents the results of the spin-boson model. In these model calculations, the temperature is $(\mathrm{k_B} T)^{-1}=5$, the energy bias is $\epsilon=1$, and the parameters for the bath are $\omega_c = 2.5$ and $\xi=0.1$. The diabatic electronic coupling is  (a) $\Delta=1$ for the adiabatic regime (such that $(\mathrm{k_B} T)^{-1}\Delta\gg 1$) or (b) $\Delta=0.1$ for the non-adiabatic regime (such that $(\mathrm{k_B} T)^{-1}\Delta\ll 1$). The results are obtained from the original diabatic SQC (open circles), QD-SQC (solid lines), and the adiabatic SQC propagation\cite{MillerAdiabatic} (dash lines), with the details of adiabatic mapping Hamiltonian provided in Appendix A. For the adiabatic SQC approach, the gradient of the derivative coupling term $\nabla d_{12}({\bf R})$ has been ignored in the dynamical propagation,\cite{MillerAdiabatic} because they are equivalent to second-order derivative couplings and very expensive to obtain in regular electronic structure calculations.\cite{MillerAdiabatic} It can be clearly seen that, while SQC and QD-SQC provide identical results (with the same numerical cost), the adiabatic SQC completely breakdown in the non-adiabatic regime presented in panel (b). This due to the fact that $\nabla d_{12}({\bf R})$ is much larger in the non-adiabatic regime (weak diabatic coupling regime) compared to the adiabatic regime (strong diabatic coupling regime). Fig.~\ref{sb-adiabatic}c and Fig.~\ref{sb-adiabatic}d depict both the first derivative coupling term $d_{12}(R)$ and its derivative $\nabla d_{12}(R)$ for a particular nuclear mode $R$ that has the closest frequency compared to $\omega_c$, with the corresponding electronic coupling in (a) and (b). One can clearly see that the derivative coupling $d_{12}(R)$ exhibit large peaks and a rapidly change near the avoiding crossing regions, which is even more pronounced for  $\nabla d_{12}(R)$, especially in the non-adiabatic regime. Thus, simply ignore it will cause large numerical error for dynamics,\cite{MillerAdiabatic} especially when it is even larger than derivative coupling itself. Comparisons between the SQC-based approaches and the numerically exact results are also provided in Appendix C. 

We should note that, with the recently developed KM-SQC approach\cite{MillerAdiabatic} (with details provided in Appendix A), the kinematic momentum transform explicitly eliminates the presence of $\nabla d_{12}({\bf R})$ term in the nuclear force (instead of ignoring it), thus help achieving accurate results in the non-adiabatic regime presented in panel (b). However, KM-SQC explicitly contains the derivative couplings in the mapping equations (see Eqn.~\ref{eqn:kmsqc} in Appendix A). Thus, it might exhibit numerical challenge when  derivative couplings are highly peaked and requires a much smaller time step for a stable propagation. 

\begin{figure}
 \centering
  \begin{minipage}[h]{\linewidth}
     \centering
     \includegraphics[width=1.0\linewidth]{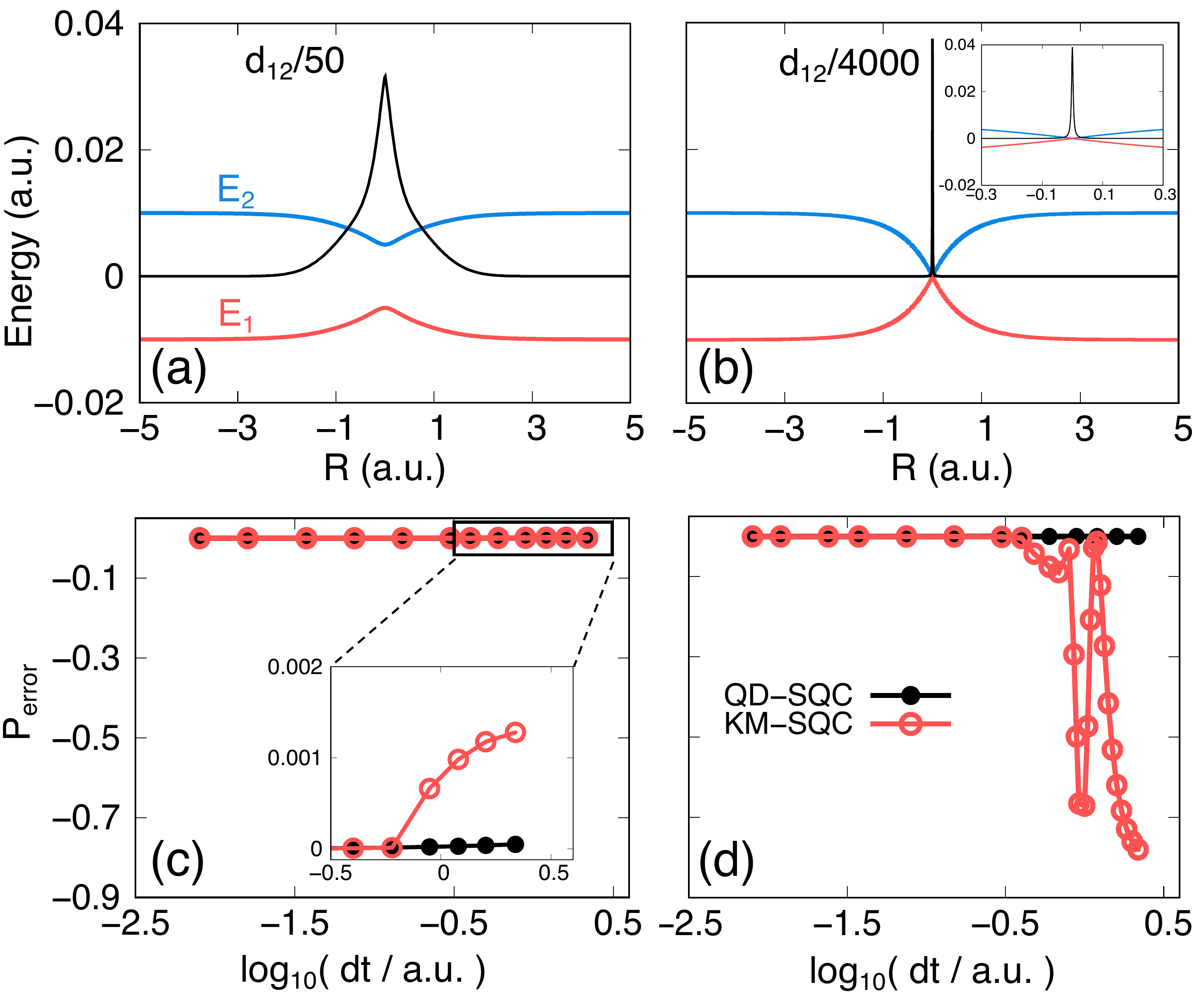}
  \end{minipage}%
   \caption{The relative error of the action variable in simple avoided crossing models. The adiabatic potentials (red and blue) and the derivative couplings (black) for the model with (a) strong and (b) weak avoid crossing are presented. The corresponding relative error are presented in (c) and (d), obtained from KM-SQC (open circle) and QD-SQC (filled circle) propagation schemes.}
\label{fig:error}
\end{figure}

Fig.~\ref{fig:error} presents the relative error of the action variable $n_j= \frac{1}{2}\left(p_j^2 + q_j^2 - 2\gamma \right)$ associated with the diabatic electronic state $|j\rangle$ at a long time ($t\rightarrow \infty$), under various nuclear time step $dt$ used for the dynamics propagation. To demonstrate the performance of various propagation schemes, here we use the original Tully's Model I (single avoided crossing model\cite{Tully}), as well as a modified version of it which contains a much narrower derivative coupling ({\it i.e.}, a weak avoid crossing model). The Hamiltonians of these two models and the corresponding parameters are provided in Appendix D. The adiabatic potentials and derivative couplings are presented in panel (a) and (b). In these simple avoid crossing models, the long-time population plateaus at a given value even at the single-trajectory level, allowing us to conveniently assess the numerical error generated from various propagation schemes. The relative error is defined as $P_\mathrm{error}=[n_{j}(dt)-n_{j}(dt\rightarrow 0)]/n_{j}(dt\rightarrow 0)$, where $n_j (dt)$ is the action obtained with a nuclear time step $dt$, and $n_j(dt\rightarrow 0)$ is the action obtained with a very small nuclear time step, such that the time-dependent action along a given trajectory as well as its long-time value converge. The electronic time step for integrating the mapping equation of motion, on the other hand, is chosen to be as small as required to converge the corresponding action $n_j$ at a given nuclear time step $dt$. For propagating mapping variables, no additional derivative couplings (for KM scheme) or electronic Hamiltonian matrix elements (for QD scheme) are computed; they are obtained based on simple linear interpolation schemes (such as Eqn.~\ref{eqn:interpolation}). This is consistent with most of the on-the-fly quantum dynamics propagation procedures,\cite{Rossky-Webster,tully94jcp,Meek2014,Subotnik2016} where the electronic structure calculations are performed only at various nuclear time steps and the quantities at electronic time steps are interpolated. This is a compromise in order to address the expensive numerical cost of expensive electronic structure calculations. The relative error is computed from a single SQC trajectory, with initial nuclear condition $R_0=-9.0$ a.u. and $P_0 = 30.$ a.u., and initial mapping condition $n_1=1.0$, $n_2=0.0$, and $\theta_1=\theta_2=\pi/4$; consistent numerical behaviors of the error with other initial conditions are also observed.  

Fig.~\ref{fig:error}c-d presents the relative errors of the action variable obtained from KM-SQC (red open circles) and QD-SQC (black filled circles). In Fig.~\ref{fig:error}c it is clear that for the model system presented in panel (a), both QD-SQC and KM-SQC provide stable propagations, generating very small numerical error even with a relatively large nuclear time step $dt$. This is because that the model in Fig.~\ref{fig:error}a has a broad derivative coupling, such that it does not change significantly on a typical time-scale that the nucleus moves. Under this scenario where the derivative coupling is well-behaved, the QD scheme does not significant numerical advantage compared to the adiabatic propagation scheme that directly uses derivative couplings.
\begin{figure}
 \centering
  \begin{minipage}[h]{\linewidth}
     \centering
     \includegraphics[width=\linewidth]{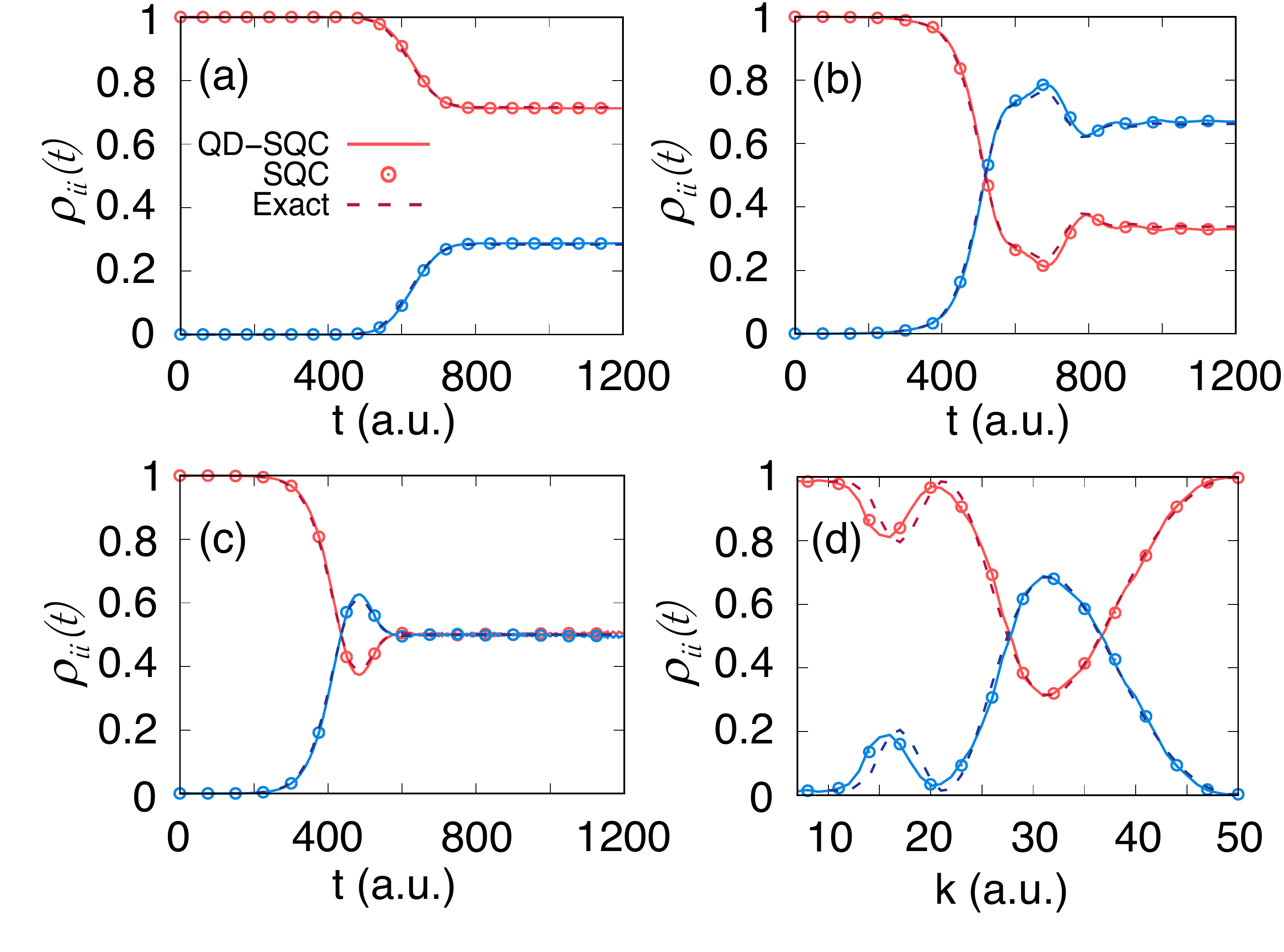}
  \end{minipage}%
   \caption{Diabatic state population of Tully's scattering models, with (a) Model I, (b) Model II, and (c) Model III. Results are obtained from SQC (open circles), QD-SQC (solid) and numerical exact calculations (dash). (d) Asymptotic diabatic population of model II as a function of various center momenta $P_0=\hbar k$ of the initial nuclear wavepacket.} 
\label{tully}
\end{figure}

In Fig.~\ref{fig:error}d, however, there is a large difference between the numerical error generated from KM-SQC and QD-SQC, especially when a large $dt$ is used. This is because that the system presented in panel (b) has a very narrow derivative coupling, such that it can spikes on a time-scale that is shorter than the nuclear time step $dt$. When using large $dt$ in KM-SQC, the nuclear position can step on different values of the derivative coupling spike or even completely step-over it and miss it,\cite{Meek2014} resulting different long time populations and an oscillatory behavior of errors. The details of the nuclear positions along a trajectory with various $dt$ are presented in Appendix D, clearly demonstrating the above mentioned behavior. We emphasize that even thought the error defined from long-time action value seems to be reduced with some larger $dt$, the overall time-dependent action variable, especially the value at the avoid crossing region are erroneous. Thus, the approaches that explicitly require derivative couplings (and use a simple linear interpolation scheme for obtaining them, as here we implemented for the KM scheme) either encounter numerical challenges or start to accumulate numerical errors.\cite{Meek2014} The QD scheme, on the other hand, provides more accurate results even when using a relatively larger $dt$, simply because that the QD schemes only requires the well-behaved transformation matrix elements $\langle \Phi_1({\bf R}(t_1))|\Phi_2({\bf R}(t_2))\rangle$ instead of the highly peaked derivative coupling ${\bf d}_{12}({\bf R})$. That being said, there might be good alternative approaches to achieve the same attractive features for dynamics propagation, such as those recently developed norm-preserving interpolation schemes.\cite{Meek2014, Subotnik2016} The QD scheme is perhaps, still the most straightforward one that allows robust dynamical propagation and enables a seamless interface between the diabatic quantum dynamics approach (such as SQC) and adiabatic electronic structure calculations. 

The extreme scenarios are systems with trivial crossings or conical intersections, where the derivative couplings become singular. Under these circumstances, the QD scheme becomes more appealing compared to the other schemes that explicitly requires derivative couplings (regardless of the detailed interpolation schemes), as they might encounter intrinsic difficulties no matter how small the $dt$ is used, due to the diverging derivative couplings. Under these circumstances, the QD scheme still provides more {\it robust} propagation of the dynamics regardless of the shape of derivative couplings, simply because it does not use the information of derivative couplings. In real molecular systems, weak avoid crossings, trivial crossings, or conical intersections are commonly encountered, making the QD scheme appealing due to its robustness. 

\begin{figure}
 \centering
  \begin{minipage}[h]{\linewidth}
     \centering
     \includegraphics[width=\linewidth]{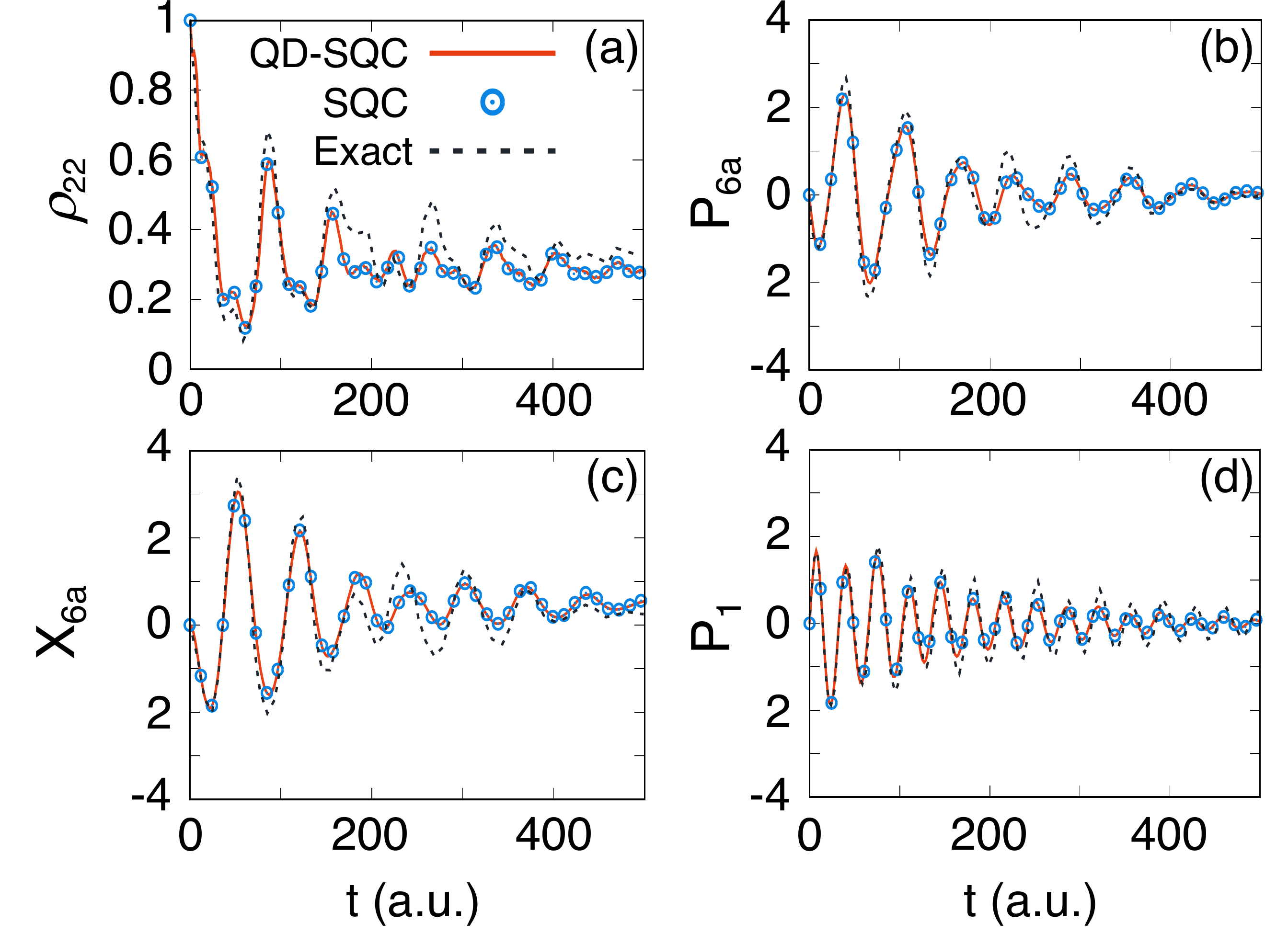}
  \end{minipage}%
   \caption{Quantum dynamics in a conical intersection model of pyrazine, with (a) diabatic population of state $|2\rangle$, (b) average momentum of the $6a$ mode, (c) average position of the 6a mode and (d) average momentum of the symmetric mode. Results are obtained from SQC (open circles), QD-SQC (solid), and numerical exact calculations (dash).}
\label{conical}
\end{figure}

Fig.~\ref{tully} presents the results of Tully's three non-adiabatic scattering models,\cite{Tully} with (a) single avoided crossing (Tully's Model I), (b) dual avoided crossing (Tully's Model II), and (c) extended coupling with reflection (Tully's Model III). These results are obtained from the diabatic SQC (open circles), QD-SQC (solid lines), and numerical exact split-operator Fourier transform method (dash lines). Initial nuclear conditions are sampled from the Wigner transformed Gaussian wavepacket, with $\Gamma$=1 a.u., $R_{0}=-9.0$  a.u., and $P_{0}=30.$ a.u. Fig.~\ref{tully}a-c provide the population $\rho_{11}(t)$ (red) and $\rho_{22}(t)$ (blue). QD-SQC gives the same results as those obtained from diabatic SQC; both are close to the numerically exact results. Fig.~\ref{tully}d presents the asymptotic diabatic population of Tully's model II as a function of the center momenta $P_{0}=\hbar k$ for the initial nuclear wavepacket. Again, QD-SQC provides the same results as the diabatic SQC, and both are close to the numerical exact ones. 

Fig.~\ref{conical} presents the results for a two-state, three-mode conical intersection model.\cite{StockConical, StockConicalJCP95} Here, the three modes are indicated as $R_{k}\in\{R_{1}$, $R_{6a}$, $R_{10a}\}$, and the model Hamiltonian has the form $\hat{H}=\sum_{k}{1 \over 2}\left[ P_k^2 + \omega_k^2 R_k^2 \right]+\sum_{i}\left[E_{i}+\sum_{k}c_{{i}_{k}}R_{k}\right]|i\rangle\langle i|+\lambda R_{10a}\left[|1\rangle \langle 2 | +|2 \rangle \langle1|\right]$. The parameters can be found in Ref. \citen{StockConical}. Both the non-adiabatic coupling element $\langle \Phi_{1}({\bf R}(t))|{{\partial}\over{\partial t}}\Phi_{2}({\bf R}(t))\rangle$ and the derivative coupling vector diverge near the conical intersection, creating numerical challenges for directly propagating dynamics in the adiabatic representation. The QD scheme avoids this challenge because it only requires $\langle \Phi_{1}({\bf R}(t_1))|\Phi_{2}({\bf R}(t_2)\rangle$ for the basis transformation during the dynamical propagation. Fig.~\ref{conical} demonstrates that QD-SQC exactly reproduces the diabatic SQC results, with (a) the diabatic population of state $|2 \rangle$ and (b)-(d) expectation values of the nuclear positions and momenta. In addition, both SQC and QD-SQC provide reasonably accurate results compared to the numerical exact ones. 
\begin{figure}
 \centering
  \begin{minipage}[t]{\linewidth}
     \centering
     \includegraphics[width=1.0\linewidth]{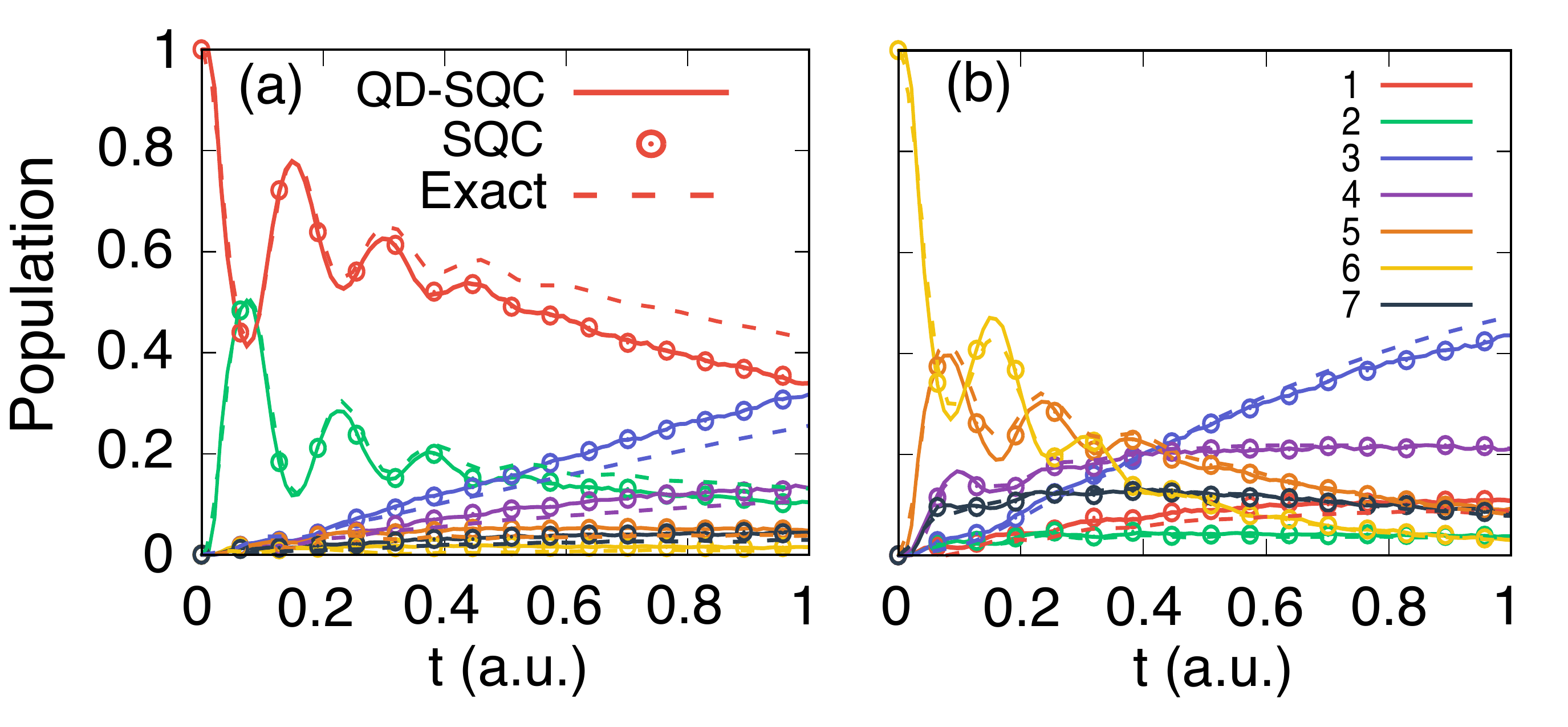}
  \end{minipage}%
   \caption{Excitation energy transfer (EET) dynamics in a model Fenna-Matthews-Olson (FMO) complex. Diabatic state populations with an initial excitation on (a) state $|1\rangle$ or (b) state $|6\rangle$ are presented. Results are obtained from SQC (open circles), QD-SQC (solid lines), and numerical exact calculations (dash lines).}
\label{fmo}
\end{figure}

Fig.~\ref{fmo} presents the quantum dynamics results for an excitation energy transfer (EET) model system.\cite{ishizaki2008pnas} Here, instead of using the triangle shaped window function, we use the original square shaped window function with a width $\gamma=0.336$. The diabatic state population are obtained from the SQC (open circles), QD-SQC (solid lines), as well as exact results (dash lines) from hierarchy equations of motion (HEOM) approach.\cite{ishizaki2008pnas} Two different initial excitation conditions are considered, with (a) state $|1\rangle$ and (b) state $|6\rangle$. As can be clearly seen, QD-SQC exactly reproduces SQC results, which are reasonably accurate compared to the numerical exact results.
\begin{figure}
 \centering
  \begin{minipage}[t]{\linewidth}
     \centering
     \includegraphics[width=0.8\linewidth]{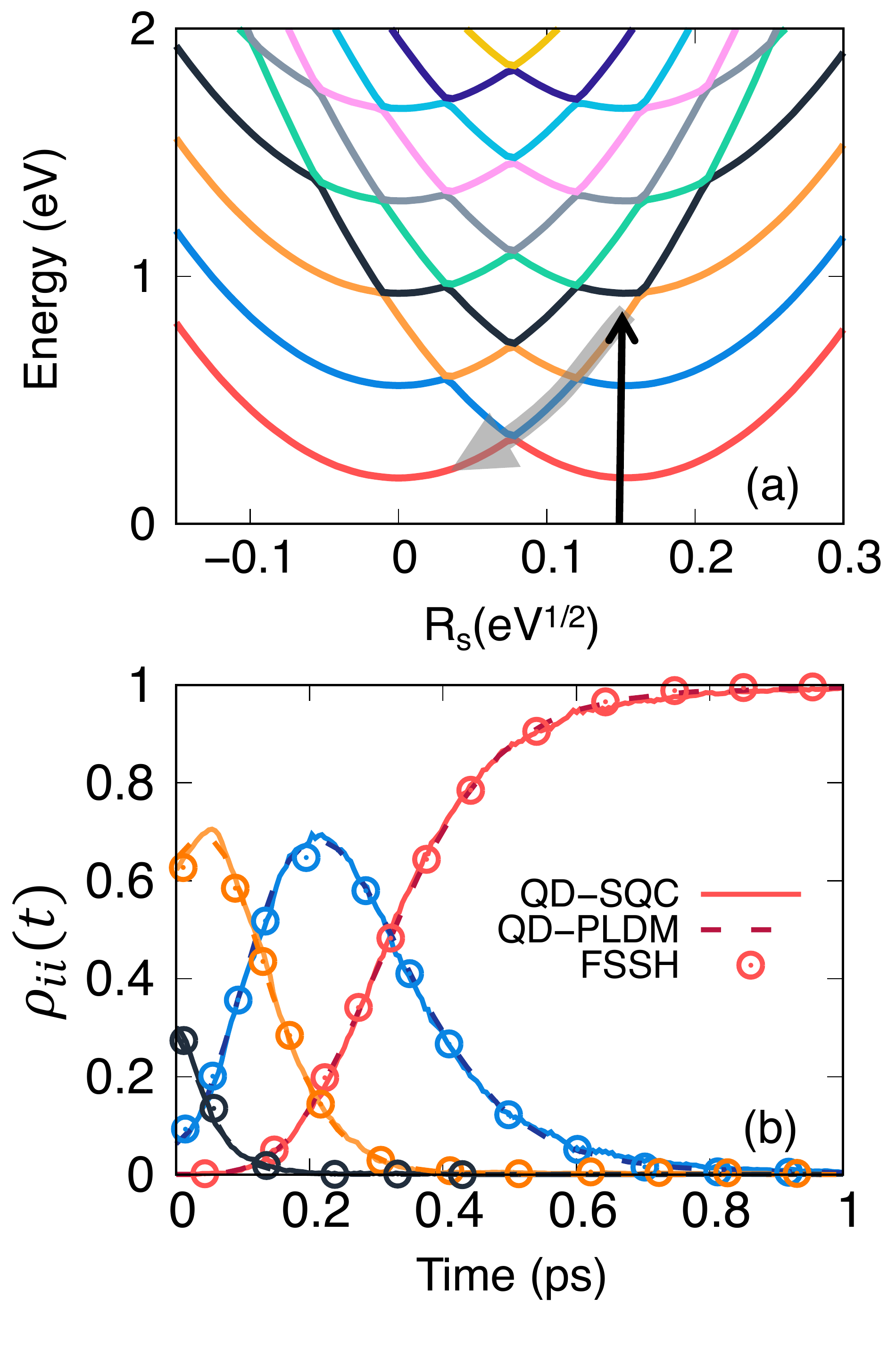}
  \end{minipage}%
   \caption{Adiabatic vibronic relaxation dynamics in a PI-PCET model system. (a) Adiabatic vibronic free energy surfaces as a function of the collective solvent coordinate. (b) The corresponding adiabatic vibronic populations obtained from QD-SQC (solid lines), QD-PLDM (dash lines), and FSSH (open circles).}
\label{fig:pipcet}
\end{figure}

Fig.~\ref{fig:pipcet} presents the {\it adiabatic vibronic} population dynamics in a PI-PCET model, with the adiabatic vibronic surfaces provided in panel (a). The initial photoexcitation is illustrated with the black arrow, and the subsequent vibrational relaxation process is illustrated with the gray arrow. The corresponding adiabatic vibronic state populations calculated using QD-SQC (solid lines) are presented in panels (b), with the same color coding used in the adiabatic vibronic potential in panel (a). Without diabatization procedure\cite{Ananth:2017,Kretchmer:2013} or using a large number of proton vibrational basis,\cite{Subotnik2018-VER,Geva2018} there is no obvious exact low-dimensional diabatic vibronic states available in this model. To assess the accuracy of the QD-SQC result, we choose to use QD-Partial Linearized Density Matrix (PLDM) path-integral approach\cite{mandal2018} (dash lines) as well as widely used FSSH approach\cite{Hazra:2010} (open circles) to simulate the same dynamical process. These two alternative approaches are proven to be accurate, at least for simulating short-time vibrational relaxation dynamics.\cite{mandal2018,Hazra:2010} The results presented in Fig.~\ref{fig:pipcet}b clearly suggests that all three approaches generate consistent dynamics, demonstrating the accuracy of the QD-SQC method. Note that the model calculations presented here with QD-SQC are different compared to the calculations with the extended-SQC.\cite{Geva2018} The latter relies on mapping the {\it strict diabatic vibrational basis} with MMST formalism, whereas QD-SQC directly uses {\it adiabatic vibronic states} to propagate dynamics. Thus, the QD propagation scheme significantly expands the scope and applicability of the SQC approach. That being said, we are aware of the recently discovered\cite{Subotnik2018-VER} convergence difficulties of SQC when including more vibrational (or vibronic) states with higher energies. Here, we explicitly avoid this issue by only including the first four adiabatic vibronic states in our QD-SQC dynamics propagation. We expect a similar issue for converging QD-SQC by including more adiabatic vibronic states; we plan to investigate this in future studies. Nevertheless, we want to emphasize that by applying the QD framework, diabatic SQC can now be directly used to propagate dynamics with adiabatic vibronic states even when there are no obvious exact low-dimensional diabatic states. 

\section{\large Conclusions}
We apply the recently developed quasi-diabatic (QD) scheme\cite{Huo2018} to propagate quantum dynamics with symmetric quasi-classical (SQC) approach.\cite{MillerJCP13} Using the instantaneous adiabatic states as the QD states during a short-time propagation, we can directly apply the diabatic SQC to propagate quantum dynamics, avoid {\it any additional} non-trivial efforts for redeveloping this approach in the adiabatic representation. The QD states are dynamically updated for each nuclear propagation step, and remain to be a convenient and compact basis for quantum dynamics propagation. In addition, the QD scheme provides a much more stable propagation compared to the adiabatic scheme as it does not explicitly require derivative couplings in the equation of motion. Further, because QD states are just the adiabatic states, they can be easily obtained from any routinely available electronic structure calculation. That being said, there might be good alternative approaches for achieving the same attractive features to propagate quantum dynamics,\cite{Meek2014, Subotnik2016} but the QD scheme is, perhaps, the simplest and the most straightforward one that allows a seamless interface between diabatic quantum dynamics approaches (such as SQC) and adiabatic electronic structure calculations. As SQC becomes an attractive approach\cite{SCFaradayMiller,Tao2016-2,Tao2016-3,Geva2018} with many appealing features,\cite{MillerJCP13,SCFaradayMiller} one of the last missing ingredients for real molecular applications is to efficiently interface it with on-the-fly electronic structure calculations. Thus, the QD-SQC approach developed in this work opens up many possibilities to perform accurate and efficient {\it ab-initio} on-the-fly simulations in complex molecular systems in future.

\section {Acknowledgement}
{This work was supported by the University of Rochester startup funds. Computing resources were provided by the Center for Integrated Research Computing (CIRC) at the University of Rochester. J. S. S appreciates valuable suggestions of the manuscript from Leopoldo Mej\'ia Restrepo. P.H. appreciates valuable discussions with Prof. Joe Subotnik and Prof. Artur Izmaylov.}

\section{Appendix A: Adiabatic MMST Hamiltonian and Kinematic Momentum Transformation}\label{A1}
Here we provide the detailed expression of the adiabatic MMST Hamiltonian. In the adiabatic representation, the total Hamiltonian in Eqn.~\ref{eqn:totalH} is expressed as the following ``vibronic'' Hamiltonian operator (with $\hbar=1$)
\begin{align}\label{eq:Hadia}
&{\hat H} =  {\hat {\bf P}^2\over 2 M}+ \sum_{\alpha} E_{\alpha}({\bf R}) |\Phi_{\alpha}({\bf R})\rangle\langle \Phi_{\alpha}({\bf R})|\\
&-\sum_{\alpha\beta} \bigg[i {\hat{\bf P}\over M} {\bf d}_{\alpha\beta}({\bf R}) + {{{D}_{\alpha\beta}({\bf R})}\over 2M} \bigg]|\Phi_{\alpha}({\bf R})\rangle\langle \Phi_{\beta}({\bf R})|,\nonumber
\end{align}
where ${\bf d}_{\alpha\beta}({\bf R}) = \langle \Phi_{\alpha} ({\bf R})| \nabla | \Phi_{\beta}({\bf R})\rangle$ is the derivative coupling vector, ${D}_{\alpha\beta} ({\bf R})= \langle \Phi_{\alpha}({\bf R}) |\nabla^2 | \Phi_{\beta}({\bf R})\rangle$ is the second-derivative coupling, and the diagonal terms ${D}_{\alpha\alpha} ({\bf R})$ are usually referred as the Born-Oppenheimer (BO) corrections.

Note that this vibronic Hamiltonian in Eqn.~\ref{eq:Hadia} can also be written as\cite{MillerAdiabatic}
\begin{eqnarray}\label{eq:Hadia2}
{\hat H} &=&\sum_{\alpha} E_{\alpha}({\bf R}) |\Phi_{\alpha}({\bf R})\rangle\langle \Phi_{\alpha}({\bf R})| \\
&&+\sum_{\alpha\beta}{1\over 2 M}\bigg(\hat {\bf P}\delta_{\alpha\beta}-i\hbar{\bf d}_{\alpha\beta}({\bf R})\bigg)^{2} |\Phi_{\alpha}({\bf R}\rangle\langle \Phi_{\beta}({\bf R})|,\nonumber
\end{eqnarray}
where the second-derivative coupling does not explicitly appear, but will indeed arise\cite{MillerAdiabatic} through the noncommutivity between $\hat{\bf P}$ and ${\bf d}_{\alpha\beta}({\bf R})$.

Applying mapping representation $|\Phi_{\alpha}({\bf R}) \rangle\langle \Phi_{\beta}({\bf R}) | \rightarrow  {\hat a}_{\alpha}^\dagger {\hat a}_{\beta} $ for the adiabatic states of the above vibronic Hamiltonian in Eqn.~\ref{eq:Hadia2} leads to the standard adiabatic MMST Hamiltonian \cite{ananth2007,MillerAdiabatic} as follows
\begin{equation} \label{eq:hm-stand}
\hat{H}= {1\over {2M}}\bigg({\bf \hat{P}} +\sum_{\alpha\beta}\hat{q}_\alpha\hat{p}_\beta{\bf d}_{\alpha\beta} ({ \bf R} )\bigg)^{2}+{1\over2}\sum_{\alpha}E_{\alpha}({\bf R})\bigg({\hat q}^2_{\alpha}+{\hat p}^{2}_{\alpha}-2\gamma\bigg),
\end{equation}
where $\gamma=0.5$ is the ZPE of the mapping oscillator.

Replacing quantum mechanical operators with classical variables, we have the following classical Hamiltonian 
\begin{equation}  \label{eq:adiaH} 
{H}=\frac{1}{2 M} (\mathbf{P + \Delta P})^2+ \frac{1}{2}\sum_{\alpha}E_{\alpha}(\mathbf{R})\left(q_{\alpha}^2+p_{\alpha}^2-2\gamma\right),
\end{equation}
where $\Delta \mathbf{P} ({\bf R},{\bf p},{\bf q}) = \sum_{\alpha\beta} q_\alpha p_\beta \mathbf{d}_{\alpha\beta}(\mathbf{R})$.
Classical equation of motion can thus be generated from the above Hamiltonian. However, it is computationally inconvenient, as the nuclear gradient explicitly dependents upon the derivative of the derivative coupling, $\nabla{\bf d}_{\alpha\beta}({\bf R})=\partial {\bf d}_{\alpha\beta}({\bf R})/\partial{\bf R}$. Evaluating this term with electronic structure calculations is equivalent to compute the second derivative couplings, which remains extremely expensive. Thus, the MMST theory in the adiabatic representation significantly increases the complexity for quantum dynamics propagations. 

In order to avoid the presence of $\nabla{\bf d}_{\alpha\beta}({\bf R})$ in the equation of motion, Cotton and Miller\cite{MillerAdiabatic} developed  the kinematic momentum (KM) transformation approach. The \textit{kinematic} momentum $\tilde{\bf P}$ is obtained through the following transformation
\begin{equation}
    \tilde{\bf P}=\mathbf{P} + \mathbf{\Delta P}.
\end{equation}
With this new set of the canonical variables, $\{{\bf R},\tilde{\bf P} \}$, one can generate an equivalent set of EOMs as follows 
\begin{eqnarray}\label{eqn:kmsqc}
    \dot{q_\alpha}&=& \frac{\partial V_\mathrm{ad}}{\partial p_\alpha}+ \sum_\beta q_\beta \mathbf{d}_{\beta\alpha}(\mathbf{R}) \cdot  \frac{\tilde{\bf P}}{M} \nonumber \\
    \dot{p_\alpha}&=&-\frac{\partial V_\mathrm{ad}}{\partial q_\alpha} + \sum_\beta p_\beta \mathbf{d}_{\beta\alpha}(\mathbf{R})\cdot \frac{\tilde{\bf P}}{M} \nonumber\\
    \dot{\mathbf{R}}&=& \frac{\mathbf{\tilde{P}}} {M}\\
    \dot{\tilde{\bf P}}&=& -\frac{\partial V_\mathrm{ad}}{\partial \mathbf{R}} +\sum_{\alpha\beta}\left (\dot{q}_{\alpha}p_\beta+q_\alpha\dot{p}_{\beta}\right)\mathbf{d}
_{\alpha\beta}(\mathbf{R}).  \nonumber  
\end{eqnarray}
Here, the adiabatic potential is defined as $V_\mathrm{ad}\left(\mathbf{q},\mathbf{p},\mathbf{R}\right) = \frac{1}{2}\sum_\alpha \left(p^2_\alpha + q^2_\alpha - 2 \gamma \right)E_\alpha\left(\mathbf{R}\right)$. On the other hand, the EOMs explicitly contain $\mathbf{d}_{\beta\alpha}(\mathbf{R})$, which could lead to numerical instabilities when these derivative couplings are highly peaked.

\section{Appendix B: Solvent parameters for the PI-PCET Model}
We provide the details of the parameters used in the PI-PCET model. The force constant for the collective solvent DOF (so-called ``inverse Pekar factor") is $f_0=4\pi \epsilon_0 \epsilon_{\infty}/(\epsilon_0 - \epsilon_{\infty})$, where $\epsilon_0$ and $\epsilon_{\infty}$ are the inertial and optical dielectric constants characterizing the polarizability of the solvent. Here, we chose these parameters that correspond to water as the solvent.\cite{Hazra:2011} 

Further, $\tau_{\text{L}}=\epsilon_{\infty}(\tau_0+\tau_{\text{D}})/\epsilon_0$ is the longitudinal relaxation time accounting for the long-time solvent response function, where $\tau_{\text{D}}$ is the Debye relaxation time and $\tau_0$ is the characteristic rotational time of the solvent molecules. All of the parameters used in this paper are tabulated in Table I and a full description of them could be found in Ref.~\citenum{Hazra:2011}.
\begin{table}[htbp]\label{parameter}
\caption{Parameters used in Langevine dynamics.}
\begin{tabular*}{0.35\textwidth}{@{\extracolsep{\fill} } l c }
\hline
{\small Parameter} & {\small Water at 298 K} \\
\hline
$\epsilon_0$ & $79.2$ \\
$\epsilon_{\infty}$ & $4.2$ \\
$f_0$ & $55.7$ \\
$\tau_0$ (ps) & $0.0103$ \\
$\tau_{\text{D}}$ (ps) & $8.72$ \\
$M_s$ ($\text{ps}^2$) & $0.265$ \\
$\lambda$ (eV) & $0.65$ \\
\hline
\end{tabular*}
\end{table}

\section{Appendix C: QD-SQC results for Spin-Boson system}
\begin{figure}
 \centering
  \begin{minipage}[h]{\linewidth}
     \centering
     \includegraphics[width=\linewidth]{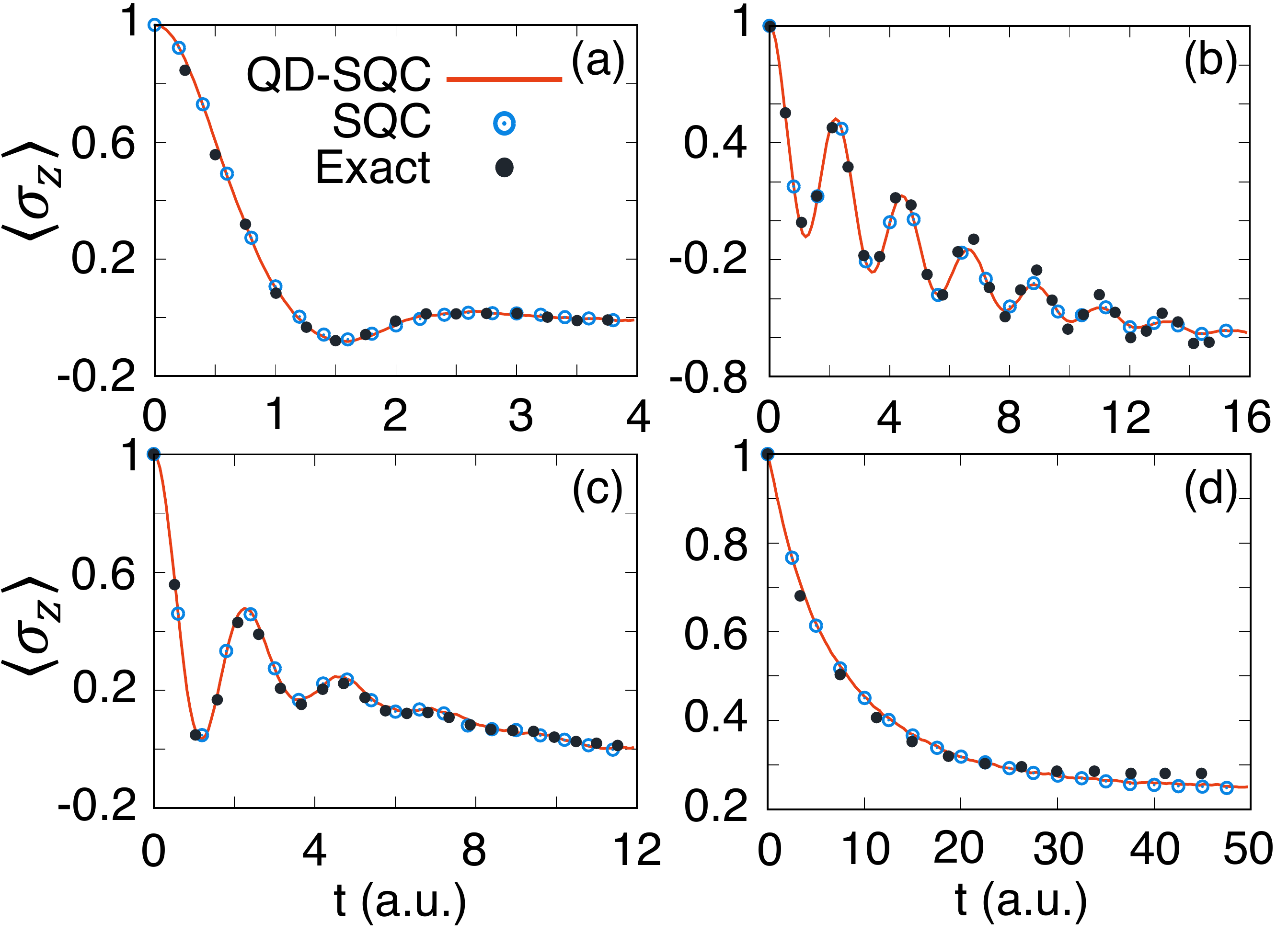}
  \end{minipage}%
   \caption{\footnotesize Population dynamics of the spin-boson model with electronic coupling $\Delta=1$ and various bias $\epsilon$ and temperature with (a) $\epsilon = 0$, $(k_\mathrm{B}T)^{-1}$ = 0.1, $\xi = 0.09$ (b) $\epsilon = 1$, $(k_\mathrm{B}T)^{-1}$= 5, $\xi = 0.25$, (c) $\epsilon = 1$, $(k_\mathrm{B}T)^{-1}$ = 0.25, $\xi = 0.1$, and (d) $\epsilon = 5$, $(k_\mathrm{B}T)^{-1}$ = 0.1, $\xi = 0.4$. Results are obtained from SQC (open circles), QD-SQC (solid lines), and numerical exact calculations (filled circles).}
\label{spin}
\end{figure}

Here, we provide additional results of spin-boson model calculations with various electronic biases and temperatures, compared to the numerical exact results. Fig.~\ref{spin} presents the population dynamics obtained from SQC, QD-SQC, and numerical exact quasi-adiabatic propagator path integral (QUAPI) calculations.\cite{Walters2016JCP,makarov1994QUAPI} In all test cases, QD-SQC (solid lines) provides identical results compared to SQC (open circles), which are close to the exact QUAPI results (filled circles).

\section{Appendix D: Model potential used in Fig.~\ref{fig:error}}
Here, we provide the Hamiltonian of the Tully's Model I used in Fig.~\ref{fig:error}, which has the following form
\begin{eqnarray}
V_{11}(R) &=& A(1-e^{-BR})~~~(\mathrm{for}~~R>0) \nonumber\\
V_{11}(R) &=& -A(1-e^{BR})~~~(\mathrm{for}~~R<0) \nonumber\\
V_{22}(R) &=& -V_{11}(R) \nonumber\\
V_{12}(R) &=& V_{21}(R) = Ce^{-DR^2} 
\end{eqnarray}
The Mass of the nuclear DOF is $M=2000$ a.u. The parameters of the potential (in a.u.) for the both models are tabulated in the following table. 
\begin{table}[htbp]
\caption{Parameters for the models presented in Fig.~\ref{fig:error}.}
\begin{tabular}{ c c c c c c } 
\hline
Parameter & $A$ & $B$ & $C$ & $D$\\
\hline
Model 1 & 0.01~~~&1.6~~~&$5\times10^{-3}$~~~&1.0~~~\\
Model 2 & 0.01~~~&1.6~~~&$5\times10^{-5}$~~~&1.0~~~\\
\hline
\end{tabular}
\end{table}

In model 2 presented in Fig.~\ref{fig:error}(b), all of the parameters are the same as in model 1 ({\it i.e.}, the original\cite{Tully} Tully's Model I), except that the diabatic coupling (modeled by parameter $C$) is reduced by 100 times, resulting a very weak avoid crossing and a highly spiked derivative coupling as depicted in Fig.~\ref{fig:error}(b).

\begin{figure}
 \centering
  \begin{minipage}[h]{\linewidth}
     \centering
     \includegraphics[width=1.0\linewidth]{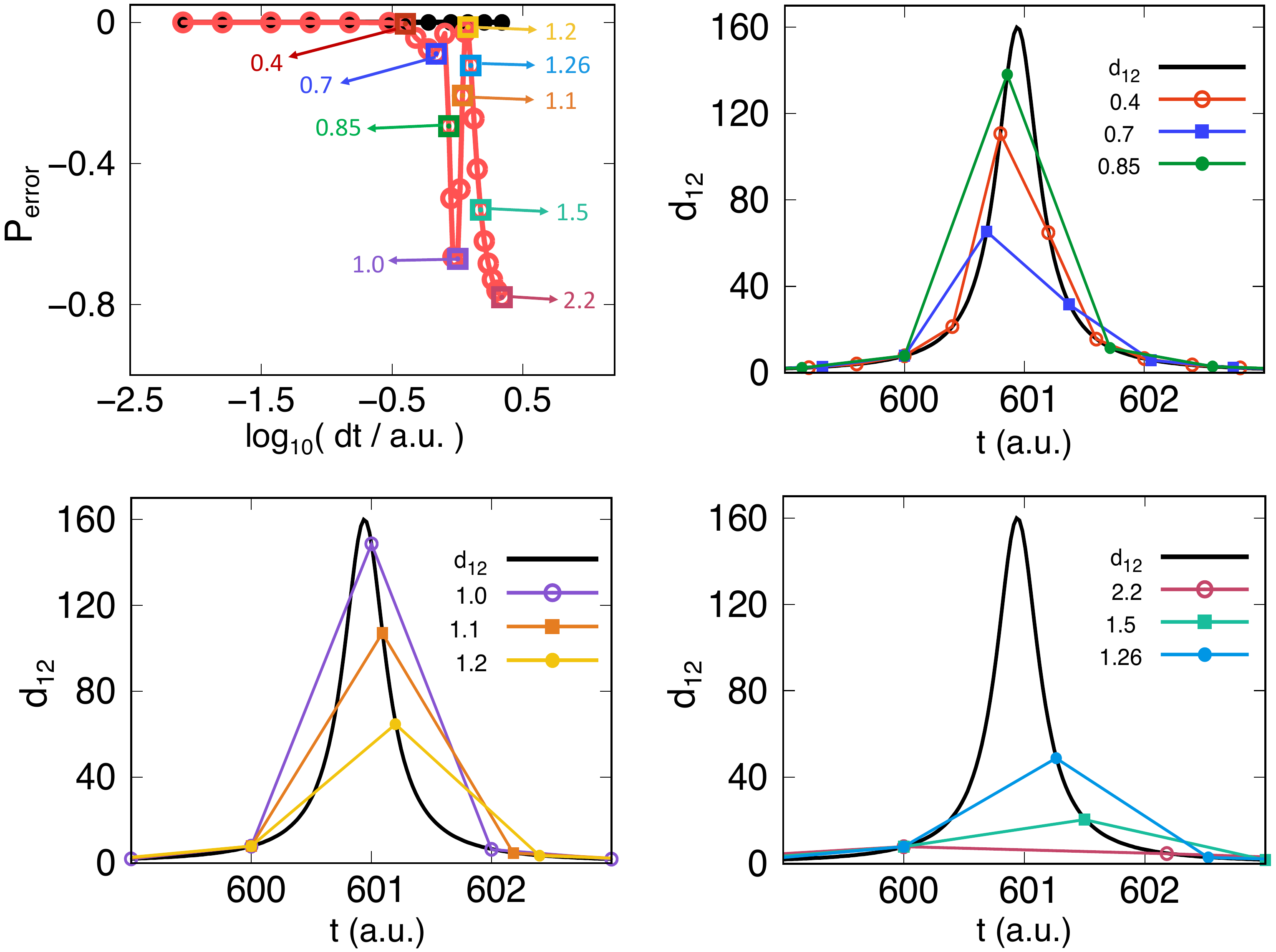}
  \end{minipage}%
   \caption{(a) The relative error of the action variable in model 2, obtained from QD-SQC (black filled circles) and KM-SQC (red filled circles). Several $dt$ (in a.u.) used in the propagation are highlighted with open squares.  (b)-(d) The derivative coupling (black) of model 2 and the nuclear position propagated with KM-SQC method with various nuclear time step. These nuclear position are presented during a time interval that the trajectory pass over the derivative coupling region, with the same color coding of the $dt$ used in panel (a).}
\label{fig:errorana}
\end{figure}

Further, in Fig.~\ref{fig:errorana} we present the derivative coupling (black) of model 2 and the nuclear position propagated with the KM-SQC approach with various nuclear time step. In panel (a), the relative error with different $dt$ are shown (same as Fig.~\ref{fig:error}(d)). In panel (b)-(d), one can clearly see that when various $dt$ is used, the nuclear position can either step on or step over the spike of the derivative coupling, resulting in a large numerical error for interpolating derivative coupling when a linear scheme is used.\cite{Meek2014} That being said, the recently developed norm preserving interpolation scheme\cite{Meek2014,Subotnik2016} can significantly reduce the numerical error compared to the simple linear interpolation scheme.\cite{tully94jcp} The QD scheme, on the other hand, explicitly avoid this issue by using the overlap integrals instead of the derivative couplings,\cite{Huo2018} as discussed in Section~\ref{sec:QD}.

\bibliographystyle{aipnum4-1}

\end{document}